\documentclass{agujournal2019}
\usepackage{url}
\usepackage{soul}
\usepackage{epstopdf, epsfig}
\usepackage{amsmath, amssymb, mathtools, gensymb}

\allowdisplaybreaks

\journalname{JGR: Space Physics}

\begin{document}


\title{Two-fluid treatment of whistling behaviour and the warm Appleton-Hartree extension}
\authors{J. De Jonghe\affil{1}\thanks{ORCID 0000-0003-2443-3903} and R. Keppens\affil{1}\thanks{ORCID 0000-0003-3544-2733}}
\affiliation{1}{Centre for mathematical Plasma-Astrophysics, KU Leuven, 3001 Leuven, Belgium.}
\correspondingauthor{Jordi De Jonghe}{jordi.dejonghe@kuleuven.be}

\begin{keypoints}
\item Group speeds can be computed numerically, highlighting whistling behaviour.
\item Whistler group speed approximations are extended to non-parallel propagation with respect to the background magnetic field.
\item The widely used Appleton-Hartree equation is extended to include a non-zero thermal electron velocity.
\end{keypoints}

\begin{abstract}
As an application of the completely general, ideal two-fluid analysis of waves in a warm ion-electron plasma, where six unique wave pair labels (S, A, F, M, O, and X) were identified, we here connect to the vast body of literature on whistler waves. We show that all six mode pairs can demonstrate whistling behaviour, when we allow for whistling of both descending and ascending frequency types, and when we study the more general case of oblique propagation to the background magnetic field. We show how the general theory recovers all known approximate group speed expressions for both classical whistlers and ion cyclotron whistlers, which we here extend to include ion contributions and deviations from parallel propagation. At oblique angles and at perpendicular propagation, whistlers are investigated using exact numerical evaluations of the two-fluid dispersion relation and their group speeds under Earth's magnetosphere conditions. This approach allows for a complete overview of all whistling behaviour and we quantify the typical frequency ranges where they must be observable. We use the generality of the theory to show that pair plasmas in pulsar magnetospheres also feature whistling behaviour, although not of the classical type at parallel propagation. Whistling of the high frequency modes is discussed as well, and we give the extension of the Appleton-Hartree relation for cold plasmas, to include the effect of a non-zero thermal electron velocity. We use it to quantify the Faraday rotation effect at all angles, and compare its predictions between the cold and warm Appleton-Hartree equation.
\end{abstract}

\section{Introduction}
The detection of whistler waves was first recorded in 1918 by radio operators who observed audio signals with a rapidly changing pitch \cite{Barkhausen1919}. Presently, they are covered in various plasma physics textbooks like \citeA{Stix1992}, \citeA{BaumjohannTreumann1997}, \citeA{TreumannBaumjohann1997}, \citeA{Bittencourt2004}, \citeA{GurnettBhattacharjee2005}, \citeA{Bellan2006}, and \citeA{ThorneBlandford2017}. In the literature these so-called whistlers are described as modes whose group speed varies drastically for small changes in frequency in the interval between the ion and electron cyclotron frequencies. This dispersive nature is observed by the aforementioned change in pitch due to waves with higher frequencies arriving first. For Earth's magnetosphere, the frequency range of this behaviour overlaps with the audible frequencies, such that the signal can be converted to a whistling sound. After their initial discovery, it was shown by \citeA{Storey1953} that the detected signals were created by lightning in the southern hemisphere and that the waves travelled along the magnetic field to the northern hemisphere. Hence, the discussion of these waves in the literature is often limited to propagation parallel to the magnetic field. However, our recent, rigorous analysis of the general, ideal, warm ion-electron dispersion relation \cite{DeJongheKeppens2020} showed that parallel propagation is uniquely special for mode behaviour in a two-fluid plasma, since it shows different connectivity between small and large wavenumber behaviour of the 6 pairs of supported modes (S, A, F, M, O, and X). The appearance of avoided crossings suggests that whistler behaviour at oblique angles differs from parallel propagation.

Later on, a whistler with frequencies below the ion cyclotron frequency was discovered by Earth orbiting spacecraft \cite{Gurnett1965}. These modes are called ion cyclotron whistlers and they occur for frequencies approaching the ion cyclotron frequency asymptotically from below for growing wavenumber. Contrary to the classical whistlers, this results in dispersive behaviour where lower frequency waves arrive first \cite{GurnettBhattacharjee2005}. In the present discussion we consider a plasma with only a single ion species. Hence, in this paper there will be only one ion cyclotron frequency and one corresponding whistler. To allow for multiple ion cyclotron whistlers, the general two-fluid derivation as done in \citeA{GoedbloedKeppensPoedts2019} has to be extended to a similar three- or multi-fluid description with one fluid for each ion species present in the plasma.

In the recent literature, whistlers have been discussed using various approaches such as electron magnetohydrodynamics \cite{Damiano2009}, electron fluid models \cite{Zhao2017}, two-fluid models \cite{Huang2019} and kinetic descriptions \cite{GarySmith2009}. Although they are often discussed as travelling parallel to the magnetic field, oblique whistler waves were also detected \cite{Cattell2008} and prompted further studies \cite{Yoon2014, Artemyev2016, Ma2017}. Furthermore, whilst they were originally discovered in Earth's magnetosphere, spacecraft observations showed that whistler waves also occur in solar wind \cite{Narita2016} and the atmospheres of Venus \cite{PerezInvernon2017} and Jupiter \cite{Imai2018}. This paper will focus on whistling behaviour in Earth's magnetosphere at parallel, oblique, and perpendicular propagation, and how it is affected by the occurrence of avoided crossings compared to parallel propagation \cite{DeJongheKeppens2020}. This includes the regular whistlers and ion cyclotron whistlers, but also any rapid variations in group speed with small changes in frequency of other wavetypes, even when it occurs outside of the audible frequency range.

Since any whistling behaviour is governed by the variation in group speed with frequency (or wavenumber), we focus on the two-fluid group speed expression. Due to the nature of the dispersion relation, the group speed cannot be easily expressed as a function of only the frequency or the wavenumber. Therefore, approximations are used to obtain single-variable group speed expressions. In this paper such textbook approximations of the group speed are extended to include ion contributions and deviations from parallel propagation with respect to the background magnetic field. Since the study here is mainly limited to Earth's magnetosphere, it should be noted that results may differ for other environments based on the 6 regimes identified in \citeA{DeJongheKeppens2020}. Depending on the prevailing plasma parameters (temperatures, field strengths, densities), the location and number of (avoided) crossings between the 6 wave pair branches was found to differ, which is of direct consequence to whistler behaviour. Finally, we also comment on the absence of whistlers in pair plasmas, which is a statement pervading the literature on pair plasma behaviour \cite{StewartLaing1992, Iwamoto1993, GaryKarimabadi2009}.

Various studies have pointed out that whistlers are usually subjected to damping effects, both collisional and collisionless \cite{Crabtree2012}. This can lead to an upper limit on the refractive index \cite{Ma2017} or spectrum gaps \cite{Hsieh2018}. At the same time, damping effects may not be significant for some modes until after several magnetospheric reflections \cite{Bell2002}. The two-fluid model of an ideal, warm ion-electron plasma in this paper does not capture any damping effects. Though Landau damping is intrinsically absent in any two-fluid description, collisional damping can be included. However, this significantly complicates the model. Despite the lack of damping effects in our two-fluid model, it can describe the propagation of whistler waves in an ideal setting. However, here the focus is on the whistling behaviour itself rather than the wave propagation. Also the conversion of whistler waves into ``lower hybrid waves" at density striations \cite{Bamber1994, Rosenberg1998, Shao2012} is beyond the scope of this paper.

The Appleton-Hartree equation \cite{Appleton1932} on the other hand is often used in magneto-ionic theory to describe high-frequency waves in a cold ion-electron plasma neglecting ion motion. This relation is valid for frequencies above the electron plasma frequency $\omega_\mathrm{pe}$ at any angle $\theta$ between the wavevector and the background magnetic field. It was pointed out by \citeA{Keppens2019_coldei} that the collisionless relation can be obtained from the polynomial description of the cold ion-electron plasma waves by taking the (unphysical) $\mu = 0$ limit, where $\mu$ denotes $Zm_\mathrm{e}/m_\mathrm{i}$ for a plasma with charge number $Z$ and electron and ion mass $m_\mathrm{e}$ and $m_\mathrm{i}$, respectively. This results in infinitely heavy ions and thus forces them to be immobile.

Recently, \citeA{Bawaaneh2013} extended the Appleton-Hartree relation, which is presented in the majority of plasma physics textbooks, to describe the high-frequency waves of a warm ion-electron plasma. The $\mu = 0$ limit can be applied to the warm ion-electron dispersion relation discussed in \citeA{DeJongheKeppens2020} to recover this warm extension of the usual Appleton-Hartree equation. With this result, the effect of the electron thermal velocity on Faraday rotation in the Appleton-Hartree equation can be investigated.

\section{Model and conventions}
For both the whistlers and the Appleton-Hartree equation we start from a two-fluid description with ions and electrons, assuming a homogeneous background at rest and charge neutrality. The linearised equations are \cite{GoedbloedKeppensPoedts2019}
\begin{align}
&\frac{\partial n_\alpha}{\partial t} + n_{\alpha, 0} \nabla\cdot\mathbf{u}_\alpha = 0 && \\
&n_{\alpha, 0} m_\alpha \frac{\partial \mathbf{u}_\alpha}{\partial t} = -\nabla p_\alpha + q_\alpha n_{\alpha, 0} (\mathbf{E} + \mathbf{u}_\alpha \times \mathbf{B}_0) && \\
&\frac{\partial p_\alpha}{\partial t} + \gamma p_{\alpha, 0} \nabla\cdot\mathbf{u}_\alpha = 0 && \\
&\nabla \times \mathbf{E} = - \frac{\partial \mathbf{B}}{\partial t} && \nabla \cdot \mathbf{E} = -\frac{e}{\epsilon_0} (n_{\mathrm{e}} - Zn_{\mathrm{i}}) \\
&\nabla \times \mathbf{B} = \frac{1}{c^2} \frac{\partial \mathbf{E}}{\partial t} - \mu_0 e n_{\mathrm{e}, 0} (\mathbf{u}_{\mathrm{e}} - \mathbf{u}_{\mathrm{i}}) && \nabla\cdot \mathbf{B} = 0
\end{align}
where $n_\alpha$ signifies number density, $\mathbf{u}_\alpha$ velocity, and $p_\alpha$ pressure, for either ions ($\alpha = \mathrm{i}$) or electrons ($\alpha = \mathrm{e}$). Furthermore, $\mathbf{E}$ and $\mathbf{B}$ are the electric and magnetic field, respectively. Equilibrium quantities are denoted with a subscript $0$. The physical constants $q_\alpha$, $\epsilon_0$, $\mu_0$, and $c$ are the charge ($Ze$ for ions, $-e$ for electrons), vacuum permittivity, vacuum permeability, and light speed, respectively, and $\gamma$ is the ratio of specific heats.

To derive the dispersion relation, all perturbations are assumed to have plane wave solutions $\exp\left[ i \left( \mathbf{k}\cdot \mathbf{r} - \omega t \right) \right]$. The full derivation is described in \citeA{GoedbloedKeppensPoedts2019} and goes back to \citeA{DenisseDelcroix1961}. A similar approach to low-frequency waves was also used by \citeA{Stringer1963}. A recent, detailed discussion can be found in \citeA{DeJongheKeppens2020} for the warm ion-electron plasma, which forms the starting point for the current discussion. Earlier work by \citeA{Zhao2015} features a similar two-fluid model, neglecting the displacement current to focus on the three low-frequency modes. The addition of the displacement current leads to a description including the electromagnetic modes (for a total of 6 forward-backward propagating wave pairs), which is clearly demonstrated in \citeA{Kakuwa2017}. Whilst this latter research includes the displacement current, it does not take density perturbations into account and is thus incompressible. The model here incorporates both the displacement current and density perturbations, which provide corrections to the aforementioned models. Furthermore, it is fully consistent with ideal relativistic MHD and includes light waves.

To summarise, the result is a polynomial dispersion relation of sixth degree in $\bar{\omega}^2$ and of fourth degree in $\bar{k}^2$ and is detailed in \citeA{GoedbloedKeppensPoedts2019} and \citeA{DeJongheKeppens2020}. Here, the frequency $\omega$ is normalised using the combined plasma frequency $\omega_\mathrm{p} = \sqrt{\omega_\mathrm{pe}^2+\omega_\mathrm{pi}^2}$ as $\bar{\omega} = \omega/\omega_\mathrm{p}$, where $\omega_{\mathrm{pe}} = \sqrt{e^2 n_{\mathrm{e}}/\epsilon_0 m_{\mathrm{e}}}$ and $\omega_{\mathrm{pi}}  = \sqrt{Z^2 e^2 n_{\mathrm{i}}/\epsilon_0 m_{\mathrm{i}}}$ are the electron and ion plasma frequencies, respectively. Similarly, the wavenumber $k$ is normalised as $\bar{k} = k\delta$, where $\delta = c/\omega_\mathrm{p}$ is the combined skin depth. The dispersion relation also features five parameters, namely the (positively defined) normalised electron and ion cyclotron frequency $E = |\omega_\mathrm{ce}|/\omega_\mathrm{p}$ and $I = \omega_\mathrm{ci}/\omega_\mathrm{p}$, respectively, with $\omega_{\mathrm{ce}} = -eB_0/m_{\mathrm{e}}$ and $\omega_{\mathrm{ci}} = ZeB_0/m_{\mathrm{i}}$, the normalised electron and ion sound speed $v = v_\mathrm{e} / c$ and $w = v_\mathrm{i} / c$, respectively, and the propagation angle parameter $\lambda = \cos\theta$, where $\theta$ is the angle between the wavevector $\mathbf{k}$ and the background magnetic field $\mathbf{B}$. Alternatively, the parameter $\mu = Zm_\mathrm{e}/m_\mathrm{i}$ called the ratio of masses over charges can be considered instead of $I$. They are related as $I = \mu E$. We will use $I$ in the discussion on whistler waves and $\mu$ in the discussion of the Appleton-Hartree equation. This (relativistically valid) dispersion relation can describe any plasma that relies solely on these parameters.

Besides the dispersion relation, we also adopt the S, A, F, M, O, X labelling scheme presented in \citeA{DeJongheKeppens2020}, which was also used earlier in the special cases of cold pair plasmas \cite{Keppens2019_coldpair}, cold ion-electron plasmas \cite{Keppens2019_coldei}, and warm pair plasmas \cite{Keppens2019_warmpair}. This labelling scheme relies on the fixed ordering of frequencies for all wavenumbers at all oblique propagation angles with respect to the background magnetic field. It has the distinct advantage that it does not use parallel and perpendicular propagation, which were shown to be special cases, as a reference for labelling wave types. Furthermore, it connects the three lowest modes, S, A, and F, to the well-known MHD slow, Alfv\'en, and fast waves in the low-frequency limit. In this way, the description also naturally introduces the normalised sound speed $c_\mathrm{s} = v_\mathrm{s} / c = \sqrt{(\mu v^2 + w^2)/(1+\mu)}$ and relativistic normalised Alfv\'en speed $c_\mathrm{a} = v_\mathrm{a} / c = \sqrt{EI/(1+EI)}$. Additionally, the description is easily transformed to focus on the refractive index $n = ck/\omega$, as is often done for both whistlers and the Appleton-Hartree relation. Besides the labelling scheme, we also use the group speed expressions in \citeA{DeJongheKeppens2020}.

Whilst the literature usually opts for a refractive index formulation for whistlers, we employ a formulation in terms of frequency and wavenumber. In particular, the resonance cone is often used to discuss oblique propagation. This cone follows from the observation that for any frequency $\omega$ below the electron cyclotron frequency $|\omega_{ce}|$, there is an angle $\theta_\mathrm{res}$ such that $\omega = |\omega_\mathrm{ce}| \cos \theta_\mathrm{res}$. Since $|\omega_\mathrm{ce}| \cos \theta$ is a resonance and thus an asymptotic upper bound on the frequency of one of the wavetypes at a propagation angle $\theta$, the resonance cone angle is the maximal propagation angle for a given frequency of that wavetype. Hence, this angle defines a cone centered around the magnetic field line in which a wave of this type and frequency is allowed to travel \cite{GurnettBhattacharjee2005}. However, as we will show, the appearance of avoided crossings allows the wavetype in which the traditional whistler occurs to exist at any frequency at oblique angles, contrary to parallel propagation where it is bounded by $|\omega_\mathrm{ce}|$. Hence, we forego the use of the resonance cone and work with a fixed propagation angle.

Concerning all figures, modes are always represented using the following colour convention: S$-$green, A$-$red, F$-$blue, M$-$purple, O$-$cyan, and X$-$black. This convention is also used when showing related quantities of these modes, e.g. the group speed. When representing the group speed $\mathbf{v}_{\mathrm{g}}$ of waves travelling at oblique angles, solid lines represent $\mathbf{v}_{\mathrm{g}} \cdot \hat{\mathbf{n}}$ whilst dashed lines represent $\mathbf{v}_{\mathrm{g}} \cdot \hat{\mathbf{b}}$. Here, $\hat{\mathbf{n}}$ and $\hat{\mathbf{b}}$ are the unit vectors along the wavevector $\mathbf{k}$ and the background magnetic field $\mathbf{B}$, respectively. When evaluating dispersion relations, group speed expressions, and their approximations, Earth's magnetosphere parameters were taken from \citeA{GoedbloedKeppensPoedts2019}: $E \simeq 0.935$, $\mu \simeq 1/1836$, $v \simeq 1.7 \times 10^{-3}$, and $w \simeq 3.9 \times 10^{-5}$ with a plasma frequency of $\omega_\mathrm{p} \simeq 5.64\ \mathrm{MHz}$. Whilst our value of $E$ is significantly larger than the $E \simeq 0.1$ used in the two-fluid study by \citeA{Huang2019}, both values describe the same regime from those identified in \citeA{DeJongheKeppens2020} and should thus be comparable.

\section{Whistlers}\label{sec:whistlers}
In the literature the term whistler has been used for different waves showing a rapid variation in group speed within a small frequency interval. Naturally, the group speed can either increase or decrease when the frequency is increased. Both behaviours have been observed, although they occur in different frequency and/or wavenumber ranges. The first type, which we will refer to as classical whistlers, are the modes with an increasing group speed for increasing frequency such that high-frequency waves of this type travel faster. These are the waves that were first observed during the First World War as descending tones \cite{Barkhausen1919, Bittencourt2004, GurnettBhattacharjee2005, Bellan2006}. A second type, which we call ascending frequency whistlers \cite{Bittencourt2004}, features the opposite behaviour. At parallel propagation, the first and second type actually correspond to the same mode in a different wavenumber range. Finally, a third type, dubbed ion cyclotron whistlers \cite{GurnettBhattacharjee2005}, also describes decreasing group speed for increasing frequency near the ion cyclotron frequency, as the name suggests.

Since all whistling behaviour relies on the variation of the group speed with the frequency or wavenumber, writing the group speed as a function of a single variable, either the frequency or the wavenumber, is of primary interest. However, due to the high polynomial degree of the dispersion relation, this is not analytically feasible. To gain insight into the behaviour of the group speed of several modes, we will use approximations of the dispersion relation to obtain approximate single-variable group speed expressions. Whilst similar approximations can be found in the literature, they are limited to parallel propagation and/or neglect ion contributions. Here, these approximations are extended to include ion effects and describe propagation at any angle. The approximations are complemented by numerical evaluations of the mode group speeds to create a complete picture of two-fluid whistling behaviour.

Although the term whistler usually refers to the low-frequency S, A, and F modes, because their frequencies lie in the audible range, similar behaviour can also be found in the high-frequency M, O, and X modes near their respective cutoffs. However, for Earth's magnetosphere these cutoffs lie well above the audible frequencies, which approximately span $20\ \mathrm{Hz}$ to $20\ \mathrm{kHz}$ for the human ear, so it is not directly ``translated" by a receiver into a whistling sound. Nevertheless, we will also consider these high-frequency modes and refer to their rapid variations in group speed as high-frequency whistling.

\subsection{Classical whistlers}\label{sec:df-whistlers}
\begin{figure}[!htb]
\centering
\includegraphics[width=\textwidth]{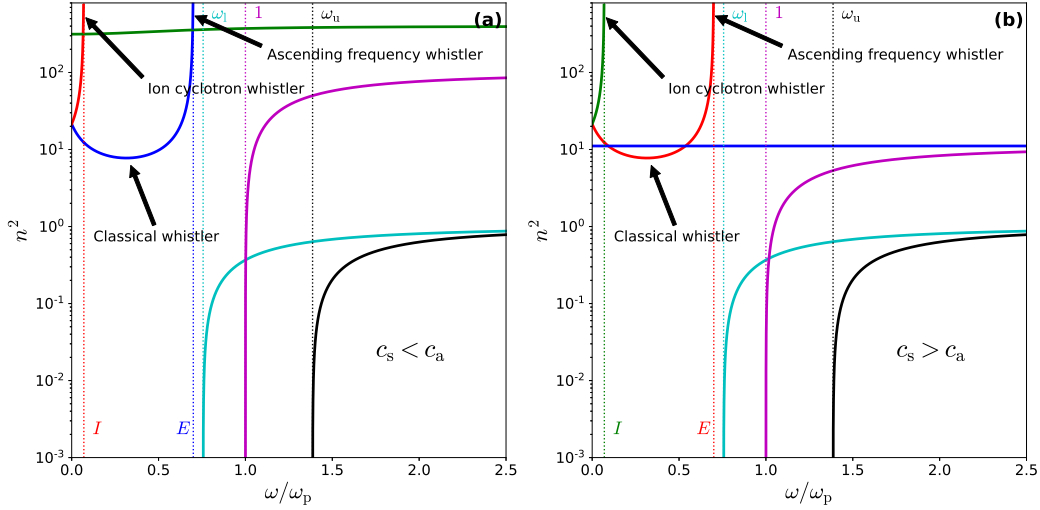}
\caption{Refractive index $n = ck/\omega$ squared as a function of the frequency at parallel propagation for the values $E = 0.7$ and $\mu = 0.1$ (values chosen for demonstrational purposes). In case (a), $v = 0.1$, $w = 0.05$, and the classical whistler mode is the (blue) F mode. In case (b), $v = w = 0.3$ and the classical whistler mode is the (red) A mode.}
\label{fig:omega-n-comparison}
\end{figure}

The classical whistlers are usually described as the whistling waves with frequencies between the ion and electron cyclotron frequency ($I$ and $E$).  Which mode that is at parallel propagation, depends on whether $c_\mathrm{s} < c_\mathrm{a}$ or $c_\mathrm{s} > c_\mathrm{a}$. In \citeA{DeJongheKeppens2020} we pointed out how up to six different parameter regimes in ($E$, $I$, $v$, $w$) can be distinguished where in each regime an a priori known amount of crossings between $\omega(k)$ branches can be identified as well as the locations of these crossings. In Fig. \ref{fig:omega-n-comparison}, we choose parameters representative of the two $E < 1$ regimes, as is the case for the magnetosphere, but similar figures can be produced for other regimes. Here, the squared index of refraction $n = ck/\omega$ is shown as a function of the frequency. In these diagrams, the mode with the classical whistler behaviour is characterised by a finite minimum index of refraction $n$ at a non-zero frequency. If $c_\mathrm{s} < c_\mathrm{a}$, as is the case for the magnetosphere, the classical whistler is related to the F mode (Fig. \ref{fig:omega-n-comparison}a, in blue). However, if $c_\mathrm{s} > c_\mathrm{a}$, it would be the A mode (Fig. \ref{fig:omega-n-comparison}b, in red).

With the use of a large index of refraction approximation ($n \gg 1$) and neglecting ion contributions, the literature offers the approximate group speed expression \cite{GurnettBhattacharjee2005}
\begin{equation}\label{eq:whistlergroup-literature}
\frac{\partial\omega}{\partial\mathbf{k}} = 2c \frac{\omega^{1/2} \left(|\omega_\mathrm{ce}| - \omega \right)^{3/2}}{|\omega_\mathrm{ce}| \omega_\mathrm{pe}}\ \hat{\mathbf{b}}, \quad \text{or equivalently,}\quad \frac{\partial\bar{\omega}}{\partial\bar{\mathbf{k}}} = 2 \frac{\bar{\omega}^{1/2} (E-\bar{\omega})^{3/2}}{E}\ \hat{\mathbf{b}},
\end{equation}
where $\hat{\mathbf{b}}$ indicates the unit vector along the background magnetic field, and the normalisation in the dimensionless form assumes $\omega_\mathrm{p} \simeq \omega_\mathrm{pe}$. This result can be obtained from the general dispersion relation, which at exactly parallel orientation splits in a quadratic and quartic factor in $\bar{\omega}^2$ \cite{DeJongheKeppens2020}, with the quartic written as
\begin{equation}\label{eq:pl-quartic}
\begin{aligned}
0 = &\bar{\omega}^{8} - \bar{\omega}^{6} \left[ 2 + E^2 + I^2 + 2 \bar{k}^{2} \right] + \bar{\omega}^{4} \left[ \left(1 + EI \right)^2 + 2 \bar{k}^{2} \left( 1 + E^2 + I^2 \right) + \bar{k}^{4} \right] \\ &- \bar{\omega}^{2} \bar{k}^2 \left[ 2 EI \left(1 + EI \right) + \bar{k}^{2} (E^2+I^2) \right] + \bar{k}^{4} E^2 I^2.
\end{aligned}
\end{equation}
This expression can be factorised further \cite{Keppens2019_coldei, DeJongheKeppens2020} to give
\begin{equation}\label{eq:pl-quartic-factored}
\begin{aligned}
&[\, \bar{\omega}^4 + \bar{\omega}^3 |E-I| - \bar{\omega}^2 (1+EI+\bar{k}^2) - \bar{\omega} \bar{k}^2 |E-I| + \bar{k}^2 EI \,] \\
\times\quad &[\, \bar{\omega}^4 - \bar{\omega}^3 |E-I| - \bar{\omega}^2 (1+EI+\bar{k}^2) + \bar{\omega} \bar{k}^2 |E-I| + \bar{k}^2 EI \,] = 0.
\end{aligned}
\end{equation}
It was pointed out in \citeA{Keppens2019_coldei} that this mixes forward-backward propagating wave pairs, and that eq. (\ref{eq:pl-quartic}) should thus be preferred. Nevertheless, if we take the second factor, only keep up to second order in $\bar{\omega}$, and ignore the ion contributions, i.e. $I = 0$ and $\omega_\mathrm{p} \simeq \omega_\mathrm{pe}$, we get the simplified dispersion relation
\begin{equation}\label{eq:whistlerdisp-literature}
0 = \bar{k}^2 (E-\bar{\omega}) - \bar{\omega}.
\end{equation}
From this expression, the literature group speed (\ref{eq:whistlergroup-literature}) is easily derived without further approximations.

\begin{figure}[!htb]
\includegraphics[width=\textwidth]{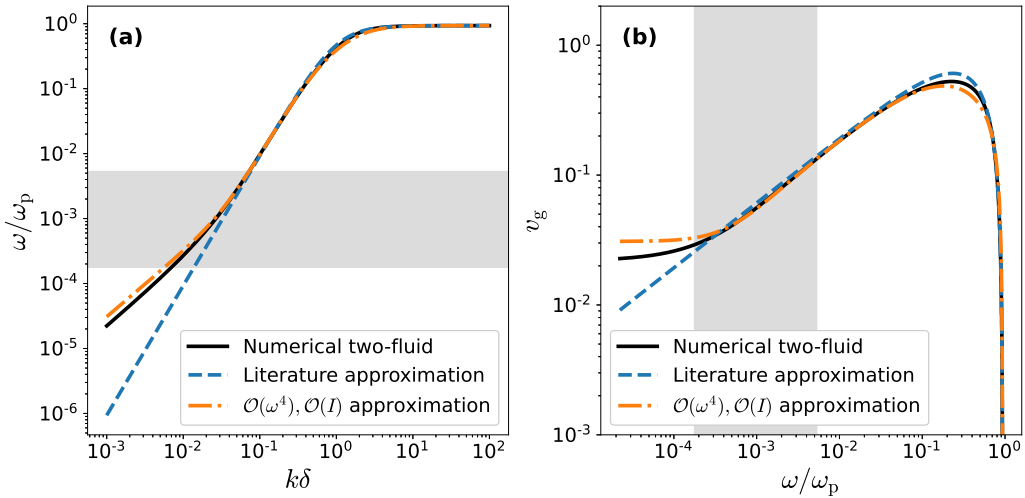}
\caption{Comparison of the literature whistler approximation and our improved approximations from eqs. (\ref{eq:whistler-plapprox}) and (\ref{eq:whistler-plgroup}) to numerical evaluations of the full two-fluid dispersion relation for Earth's magnetosphere parameters at parallel propagation. The grey areas indicate the observed frequency range shown in \citeA{GurnettBhattacharjee2005}. (a) Whistler mode dispersion diagram showing frequency variation with wavenumber. (b) Whistler mode group speed showing group speed magnitude versus frequency.}
\label{fig:ms-whistler}
\end{figure}

Clearly, keeping ion contributions in (\ref{eq:pl-quartic-factored}) results in a dispersion relation that is quadratic in $\bar{\omega}$, which is more difficult to treat. However, to expand upon the literature approximation we can start from eq. (\ref{eq:pl-quartic}) and keep the $\mathcal{O}(\bar{\omega}^4)$ and $\mathcal{O}(I)$ terms to get
\begin{equation}\label{eq:whistler-plapprox}
\bar{\omega}^2 = \frac{\bar{k}^2 (2EI+\bar{k}^2 E^2)}{1+2EI+2\bar{k}^2 (1+E^2)+\bar{k}^4}.
\end{equation}
From this relation, the phase and group speed expressions with the first order ion correction can be obtained,
\begin{equation}
\mathbf{v}_{\text{ph}} = \left[ \frac{2EI+\bar{k}^2 E^2}{1+2EI+2\bar{k}^2 (1+E^2)+\bar{k}^4} \right]^{1/2} \hat{\mathbf{b}}
\end{equation}
and
\begin{equation}\label{eq:whistler-plgroup}
\frac{\partial\bar{\omega}}{\partial\bar{\mathbf{k}}} = \frac{2\bar{k}}{\bar{\omega}} \frac{(EI+\bar{k}^2E^2)(1+2EI)+\bar{k}^4 [E^2 (1+E^2)-EI]}{[1+2EI+2\bar{k}^2 (1+E^2)+\bar{k}^4]^2}\ \hat{\mathbf{b}},
\end{equation}
where we made use of $\hat{\mathbf{b}} = \mathbf{k}/k \equiv \hat{\mathbf{n}}$ for parallel propagation. Using eq. (\ref{eq:whistler-plapprox}) to substitute $\bar{\omega}$ or $\bar{k}$ in eq. (\ref{eq:whistler-plgroup}) yields a group speed expression in one variable. How this $\mathcal{O}(I)$ approximation compares to the usual literature approximation is shown in Fig. \ref{fig:ms-whistler} along with a numerical evaluation of the full dispersion relation and an indication of observed whistler frequencies. In the classical whistler region, where the group speed is increasing with frequency $\bar{\omega}$, it follows the numerical result more closely.

\subsection{Ascending frequency whistlers}\label{sec:afw-pl}
The term ascending frequency whistler refers to the whistling behaviour that occurs near, but below, the electron cyclotron resonance $\bar{\omega} = E$ at parallel propagation \cite{Bittencourt2004}. The mode that approaches this resonance once again depends on the parameter regime. Looking at Fig. \ref{fig:omega-n-comparison} again, the ascending frequency whistler is the mode asymptotically approaching $\bar{\omega} = E$ from the left. At parallel propagation, this is the F mode if $c_\mathrm{s} < c_\mathrm{a}$ and the A mode if $c_\mathrm{s} > c_\mathrm{a}$. Note that both the classical whistler and the ascending frequency whistler are hence described by the same mode at parallel propagation, regardless of the regime. At oblique angles, the situation will be more nuanced due to the avoided crossing near the electron cyclotron frequency.

As can be seen in Fig. \ref{fig:ms-whistler}, the approximation (\ref{eq:whistlergroup-literature}) is usually used for both classical and ascending frequency whistlers. Sometimes, for ascending frequency whistlers this is approximated further by using $\omega \simeq \omega_\mathrm{ce}$ in the non-vanishing factor \cite{ThorneBlandford2017},
\begin{equation}\label{eq:whistlergroup-tb}
\frac{\partial\omega}{\partial\mathbf{k}} = \frac{2|\omega_\mathrm{ce}|c}{\omega_\mathrm{pe}} \left( 1-\frac{\omega}{|\omega_\mathrm{ce}|} \right)^{3/2} \hat{\mathbf{b}}.
\end{equation}
However, since $\bar{\omega}$ is close to $E$, the small frequency approximation used to obtain this result may no longer be the best approach. Alternatively, observe that the resonance behaviour $\bar{\omega} \rightarrow E$ occurs in the short wavelength (large $k$) limit. Hence, we will instead consider a large wavenumber approximation.

In order to expand the literature approximation to include ion effects, we only keep terms of order $\mathcal{O}(\bar{k}^2)$ or higher in eq. (\ref{eq:pl-quartic}), discard the rest, and assume that $I < \bar{\omega} < E$. This leads to a phase speed of
\begin{equation}\label{eq:whistlerphase-fullpl}
\mathbf{v}_{\mathrm{ph}} = \left[ \frac{(E^2-\bar{\omega}^2) (\bar{\omega}^2-I^2)}{-2 \left[ \bar{\omega}^4 - \bar{\omega}^2 (1+E^2+I^2) + EI (1+EI) \right]} \right]^{1/2} \hat{\mathbf{b}}
\end{equation}
and a group speed
\begin{equation}\label{eq:whistlergroup-fullpl}
\begin{aligned}
\frac{\partial\bar{\omega}}{\partial\bar{\mathbf{k}}} = &\frac{E^{3/2}}{2} \left\{ -2 \left[ \bar{\omega}^4 - \bar{\omega}^2 (1+E^2+I^2) + EI (1+EI) \right] \right\}^{1/2} (\bar{\omega}^2 -I^2)^{3/2} (\bar{\omega}+E)^{3/2} \\ &\big\{ \bar{\omega}^8 - 2\bar{\omega}^6 (E^2+I^2) + \bar{\omega}^4 \left[ (E^2+I^2)(1+E^2+I^2) - EI (1-2EI) \right] \\ &- 2\bar{\omega}^2 E^2I^2 (1+E^2+I^2) + E^3I^3 (1+EI) \big\}^{-1} \left(1 - \frac{\bar{\omega}}{E}\right)^{3/2} \hat{\mathbf{b}},
\end{aligned}
\end{equation}
where we once again used that $\hat{\mathbf{b}} = \mathbf{k}/k \equiv \hat{\mathbf{n}}$ for parallel propagation. These expressions are real if $0 < I < E$, which is guaranteed for an ion-electron plasma since $I = \mu E$ and $0 < \mu < 1$. Although we started from a different approximation, note that the group speed expression (\ref{eq:whistlergroup-fullpl}) reduces to eq. (\ref{eq:whistlergroup-tb}) if all ion terms are set to zero ($I = 0$, $\omega_\mathrm{p} = \omega_\mathrm{pe}$) and the approximation $\bar{\omega} = E$ is used in all factors except for the last one, which would vanish. However, if the ion terms are kept, using $\bar{\omega} = E$ gives an ion correction in both the phase speed,
\begin{equation}\label{eq:afwphase-ioncorr}
\mathbf{v}_{\mathrm{ph}} = \sqrt{E(E+I)} \left( 1-\frac{\bar{\omega}}{E} \right)^{1/2}\ \hat{\mathbf{b}},
\end{equation}
and the group speed,
\begin{equation}\label{eq:afwgroup-ioncorr}
\frac{\partial\bar{\omega}}{\partial\bar{\mathbf{k}}} = 2\sqrt{E(E+I)} \left( 1-\frac{\bar{\omega}}{E} \right)^{3/2} \hat{\mathbf{b}}.
\end{equation}
Note that the factor $c$ is implicit since we are working with dimensionless quantities. However, for a proton-electron plasma under magnetosphere conditions the correction introduced by the inclusion of ion contributions is negligible compared to the textbook approximation (\ref{eq:whistlergroup-tb}).

Looking at expression (\ref{eq:afwgroup-ioncorr}), the ascending frequency whistling behaviour can be seen in the last factor $(1-\bar{\omega}/E)^{3/2}$. As the frequency comes closer to the resonance, this factor becomes smaller. Since $\omega \xrightarrow{<} E$, the higher frequencies have a lower group speed. An observer listening for these signals will notice that the lower frequencies arrive first, resulting in an apparent increase in pitch.

\subsection{Ion cyclotron whistlers}\label{sec:ioncycl-whistlers}
Just like the ascending frequency whistler, the ion cyclotron whistler occurs in the resonance regime, but here the frequency approaches the ion cyclotron frequency asymptotically for increasing wavenumber, $\bar{\omega} \rightarrow I$, at parallel propagation. Once again, which mode describes these whistlers depends on whether $c_\mathrm{s} < c_\mathrm{a}$ or vice versa. If $c_\mathrm{s} < c_\mathrm{a}$, they are related to the A mode (see Fig. \ref{fig:omega-n-comparison}(a)$-$red) whereas they occur in the S mode (see Fig. \ref{fig:omega-n-comparison}(b)$-$green) if $c_\mathrm{s} > c_\mathrm{a}$.

\begin{figure}[!htb]
\includegraphics[width=\textwidth]{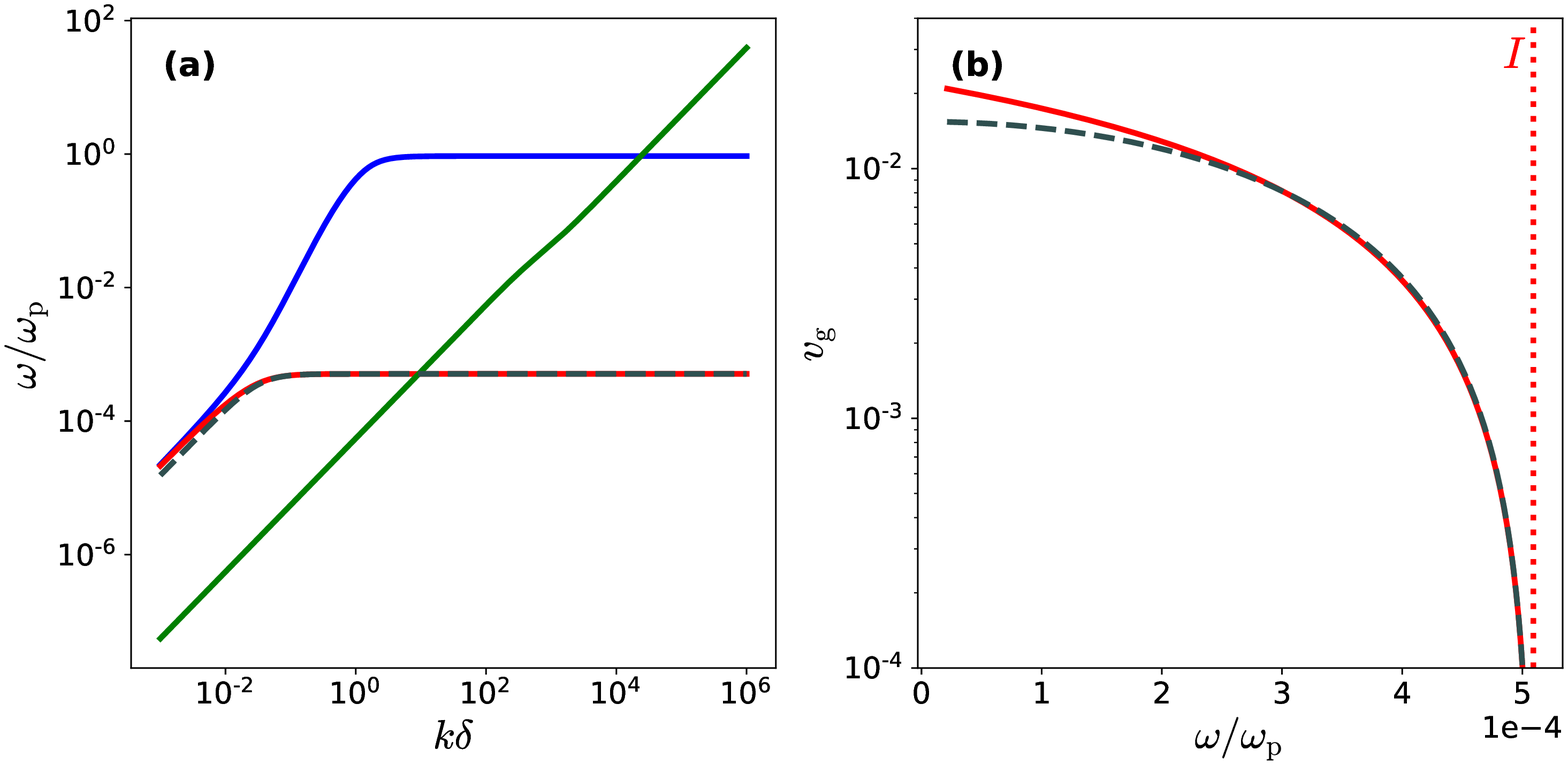}
\caption{(a) Dispersion diagram of the S, A, and F modes for Earth's magnetosphere parameters at parallel propagation. In this situation, the ion cyclotron whistler is associated with the (red) A mode. The dashed grey line is the ion cyclotron whistler approximation (\ref{eq:disp-ioncycl-pl}). (b) The group speed of the ion cyclotron whistler (A mode, red) and its approximation (\ref{eq:gs-ioncycl-pl}) (dashed grey).}
\label{fig:ioncycl-whistler}
\end{figure}

At parallel propagation, the ion cyclotron resonance is contained within the quartic branch from eq. (\ref{eq:pl-quartic}) and corresponds to the lowest frequency solution. Hence, to obtain approximate phase and group speed expressions, we start from eq. (\ref{eq:pl-quartic}) and keep only $\mathcal{O}(\bar{\omega}^2)$ terms. This gives
\begin{equation}\label{eq:disp-ioncycl-pl}
0 = \bar{\omega}^2 \left[ 2EI(1+EI) + k^2 (E^2+I^2) \right] - k^2 E^2 I^2.
\end{equation}
From this expression, the approximate phase and group speed expressions, as functions of the frequency, are
\begin{equation}
\mathbf{v}_{\mathrm{ph}} = \left[ \frac{E^2 I^2 - \bar{\omega}^2 (E^2+I^2)}{2EI(1+EI)} \right]^{1/2} \hat{\mathbf{b}}
\end{equation}
and
\begin{equation}\label{eq:gs-ioncycl-pl}
\frac{\partial\bar{\omega}}{\partial\bar{\mathbf{k}}} = \frac{\left[ E^2I^2 - \bar{\omega}^2 (E^2+I^2) \right]^{3/2}}{E^2 I^2 \left[ 2EI(1+EI) \right]^{1/2}}\ \hat{\mathbf{b}}.
\end{equation}
The exact numerical solution obtained from eq. (\ref{eq:pl-quartic}) and the approximation (\ref{eq:gs-ioncycl-pl}) are shown in Fig. \ref{fig:ioncycl-whistler}. The whistling behaviour is captured fairly well.

\subsection{High-frequency whistlers}\label{sec:hfw}
\begin{figure}[!htb]
\includegraphics[width=\textwidth]{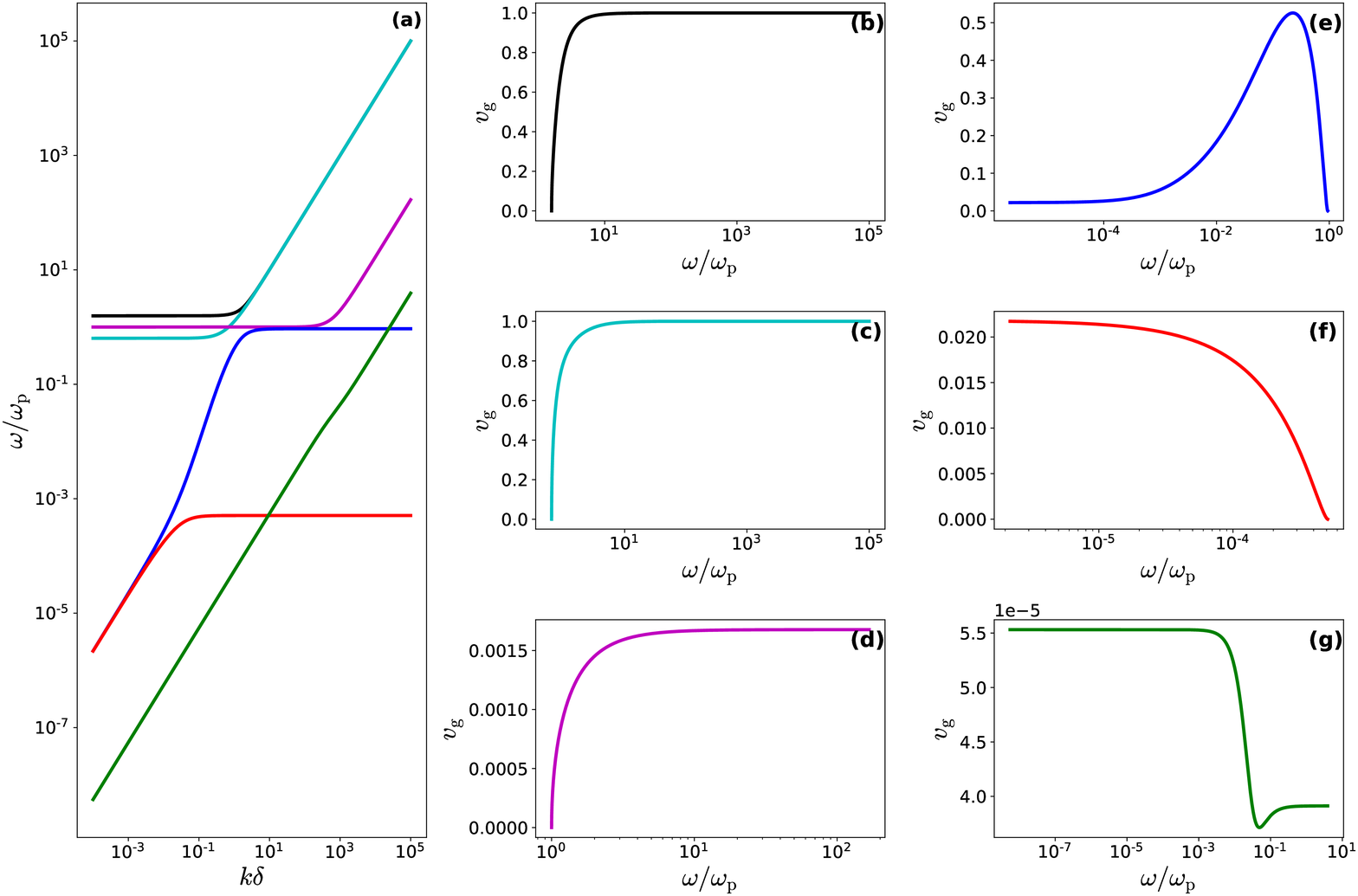}
\caption{(a) Dispersion diagram of magnetosphere conditions at parallel propagation ($E < 1$, $c_\mathrm{s} < c_\mathrm{a}$). (b-g) Mode group speeds along the magnetic field for the (b) X mode, (c) O mode, (d) M mode, (e) F mode, (f) A mode, and (g) S mode, for varying frequency ranges based on cutoffs and/or resonances.}
\label{fig:hfw}
\end{figure}

Although the high-frequency M, O, and X modes only propagate at frequencies well above the audio frequency range, the group speed of all three modes increases quickly for increasing frequency near their respective cutoffs, as can be seen for parallel propagation in Fig. \ref{fig:hfw} in panels (b), (c), and (d). This is indicative of a whistler that descends in frequency. Since these cutoff frequencies define the lower bounds for these modes, observations focused on the frequency bands slightly above each cutoff should be able to observe this behaviour. For Earth's magnetosphere, these cutoffs are approximately $\omega_\mathrm{M} = 5.64\ \mathrm{MHz}$, $\omega_\mathrm{O} = 3.59\ \mathrm{MHz}$ and $\omega_\mathrm{X} = 8.87\ \mathrm{MHz}$.

Comparing Fig. \ref{fig:hfw}(d) to (e) and (f), it should be noted that the group speed range of the M mode whistling is significantly smaller than that of both the F and A mode whistling. Additionally, the O and X mode group speeds cover the full velocity range from $0$ to $1$ (see Fig. \ref{fig:hfw}(b) and (c)), which is in full accord with their light wave behaviour. Hence, observations of the different whistling behaviours along the magnetic field would happen on different timescales, assuming the travelling distance is fixed. To demonstrate this, consider a magnetic dipole to model Earth's magnetic field and a field line at an angle of $15\degree$ to the dipole axis at a distance of $1$ Earth radius ($1\ r_\mathrm{E}$) away from the dipole. Then the length of the field line from the northern to the southern surface is approximately $l \simeq 39\ r_E$. Under a crude assumption, if the plasma parameters were to stay the same along the entire trajectory such that the travel time is simply $t(\bar{\omega}) = l/cv_\mathrm{g}(\bar{\omega})$, the classical whistler (F mode, from $\bar{\omega} = I$ up to the frequency with maximal group speed) would be observed in an interval of $\sim 20\ \mathrm{s}$, the ascending frequency whistler (F mode, from the frequency with maximal group speed up to $\bar{\omega} = 0.99\ E$) in $\sim 7\ \mathrm{min}$, the ion cyclotron whistler (A mode, from $\bar{\omega} = 10^{-5}$ up to $\bar{\omega} = 0.95\ I$) in $\sim 30\ \mathrm{min}$, the M mode whistling in the frequency interval $[1.01, 10]\ \omega_{\mathrm{p}}$ in $\sim 50\ \mathrm{min}$, the O mode whistling (from $1.01$ times the cutoff frequency up to the frequency where $v_\mathrm{g} = 0.99$) in $\sim 4\ \mathrm{s}$, and the X mode whistling (from $1.01$ times the cutoff frequency up to the frequency where $v_\mathrm{g} = 0.99$) in $\sim 0.1\ \mathrm{s}$. Therefore, different whistling behaviours occur on different timescales.

\subsection{Pair plasma whistlers}\label{sec:pair-whistlers}
As a small excursion to the cases discussed thus far, where we always used Earth's magnetosphere conditions and looked at purely parallel propagation, we here briefly consider the case of a pulsar magnetosphere. There, the strong electromagnetic fields present create a pair plasma satisfying $E = I$, which is a rather different magnetospheric environment than encountered on Earth. Although the treatment below is mostly parameter independent, we consider a typical pulsar with a period of $P = 0.5\ \mathrm{s}$ and a magnetic field strength of $B = 10^8\ \mathrm{T}$ \cite{Lyutikov1999}. Additionally, we use a sound speed of $v^2 = 0.3$ (i.e. a relativistically hot plasma that reached its maximal sound speed limit) and a Goldreich-Julian density estimate \cite{GoldreichJulian1969}, resulting in a cyclotron frequency of $E \simeq 5.89\times 10^{8}$ with a plasma frequency of $\omega_\mathrm{p} \simeq 29.8\ \mathrm{GHz}$. (We hereby correct two erronous values shown in Fig. 5b of \citeA{DeJongheKeppens2020} for the ``Pulsar Wind" and ``Magnetar Wind" cases. Their depicted number density values correspond to units of particles per $\mathrm{cm}^3$ rather than particles per $\mathrm{m}^3$, as indicated on the axis. The correct number densities are thus a factor of $10^6$ greater than indicated. Their magnetic fields are still well above the indicated critical magnetic field, so the discussion is unaffected.)

As pointed out by \citeA{StewartLaing1992}, there are no classical whistlers in equal-mass plasmas. To consider such a pair plasma in our formalism at parallel propagation, it suffices to substitute $I$ by $E$ in our dispersion relation (\ref{eq:pl-quartic}). In this case, the dispersion relation reduces to
\begin{equation}\label{eq:disp-pair}
0 = \left[ \bar{\omega}^4 - \bar{\omega}^2 (1+E^2+\bar{k}^2) + E^2 \bar{k}^2 \right]^2.
\end{equation}
It was noted by \citeA{Keppens2019_warmpair} that the warm pair dispersion relation factorises into two third order branches in $\bar{\omega}^2$, named the XFS and OMA branches. Substituting $\lambda = 1$ in these branches reveals that one factor in eq. (\ref{eq:disp-pair}) comes from the XFS branch and the other from the OMA branch. Due to this degeneracy, the phase and group speeds also simplify significantly,
\begin{align}
\mathbf{v}_{\text{ph}} &= \left( \frac{E^2-\bar{\omega}^2}{1+E^2-\bar{\omega}^2} \right)^{1/2}\ \hat{\mathbf{b}} \\
&\simeq E \left( 1-\frac{\bar{\omega}^2}{E^2} \right)^{1/2}\ \hat{\mathbf{b}} \qquad \text{for } \bar{\omega} \rightarrow E
\end{align}
and
\begin{align}
\mathbf{v}_{\text{g}} &= \frac{\bar{k} (E^2-\bar{\omega}^2)}{\bar{\omega} (1+E^2+\bar{k}^2-2\bar{\omega}^2)}\ \hat{\mathbf{b}} \\
&= \frac{(1+E^2-\bar{\omega}^2)^{1/2} (E^2-\bar{\omega}^2)^{3/2}}{E^2+(E^2-\bar{\omega}^2)^2}\ \hat{\mathbf{b}} \label{eq:group-pair} \\
&\simeq E \left( 1-\frac{\bar{\omega}^2}{E^2} \right)^{3/2}\ \hat{\mathbf{b}} \qquad \text{for } \bar{\omega} \rightarrow E. \label{eq:group-pair-approx}
\end{align}
Eq. (\ref{eq:group-pair}) is a decreasing function on the frequency interval $(0, E)$, where any classical whistler mode would be. Hence, we indeed conclude that there is no classical whistler behaviour at parallel propagation. However, near the electron cyclotron frequency $E$ the group speed decreases rapidly for increasing frequency, as is immediately clear from the approximation (\ref{eq:group-pair-approx}). Therefore, a pair plasma does have an ascending frequency whistler. In fact, due to the degeneracy in eq. (\ref{eq:disp-pair}), a pair plasma has two coinciding, indistinguishable modes with ascending frequency whistling behaviour, one associated to each of the species. Note that this mode degeneracy is lifted at oblique angles. These are the A and F modes if $c_\mathrm{s} < c_\mathrm{a}$, as is the case in a typical pulsar magnetosphere. If a pair plasma would occur in an environment such that $c_\mathrm{s} > c_\mathrm{a}$, these would be the S and A modes instead.

\subsection{Whistling at oblique angles}
Returning to the general case, and specifying parameters to the Earth's magnetosphere again, we now discuss how whistling behaviour is certainly not limited to purely parallel propagation alone. Whilst whistlers are often only considered as propagating (close to) parallel to the magnetic field, the polynomial two-fluid formalism allows us to extend the discussion to all angles. As explained in \citeA{Keppens2019_coldei, Keppens2019_coldpair, Keppens2019_warmpair, DeJongheKeppens2020}, it is impossible to make an unambiguous wave labelling scheme starting from purely parallel and purely perpendicular orientations, since those behave differently from all intermediate oblique orientations, where the wave frequencies are ordered. Due to the appearance of avoided crossings at oblique angles, which were discussed at length in \citeA{DeJongheKeppens2020}, the complete picture of whistling behaviour is more complicated. First of all, whereas in the case of parallel propagation the group speed is always along the direction of the magnetic field, as seen in expressions (\ref{eq:whistler-plgroup}), (\ref{eq:whistlergroup-fullpl}), (\ref{eq:gs-ioncycl-pl}), and (\ref{eq:group-pair}), the group speed of any wave S, A, F, M, O, or X for oblique propagation now has contributions along the directions of both the magnetic field and the wavevector. Secondly, at an avoided crossing, new whistling behaviour may appear in the two involved modes. Since the number of avoided crossings depends on the regime \cite{DeJongheKeppens2020}, the whistling behaviour at oblique angles also depends on this regime. In this section we focus on the parameter regime that is representative of the Earth's magnetosphere, $E < 1$ and $c_\mathrm{s} < c_\mathrm{a}$, using the magnetosphere parameters from \citeA{GoedbloedKeppensPoedts2019}.

\begin{figure}[!htb]
\includegraphics[width=\textwidth]{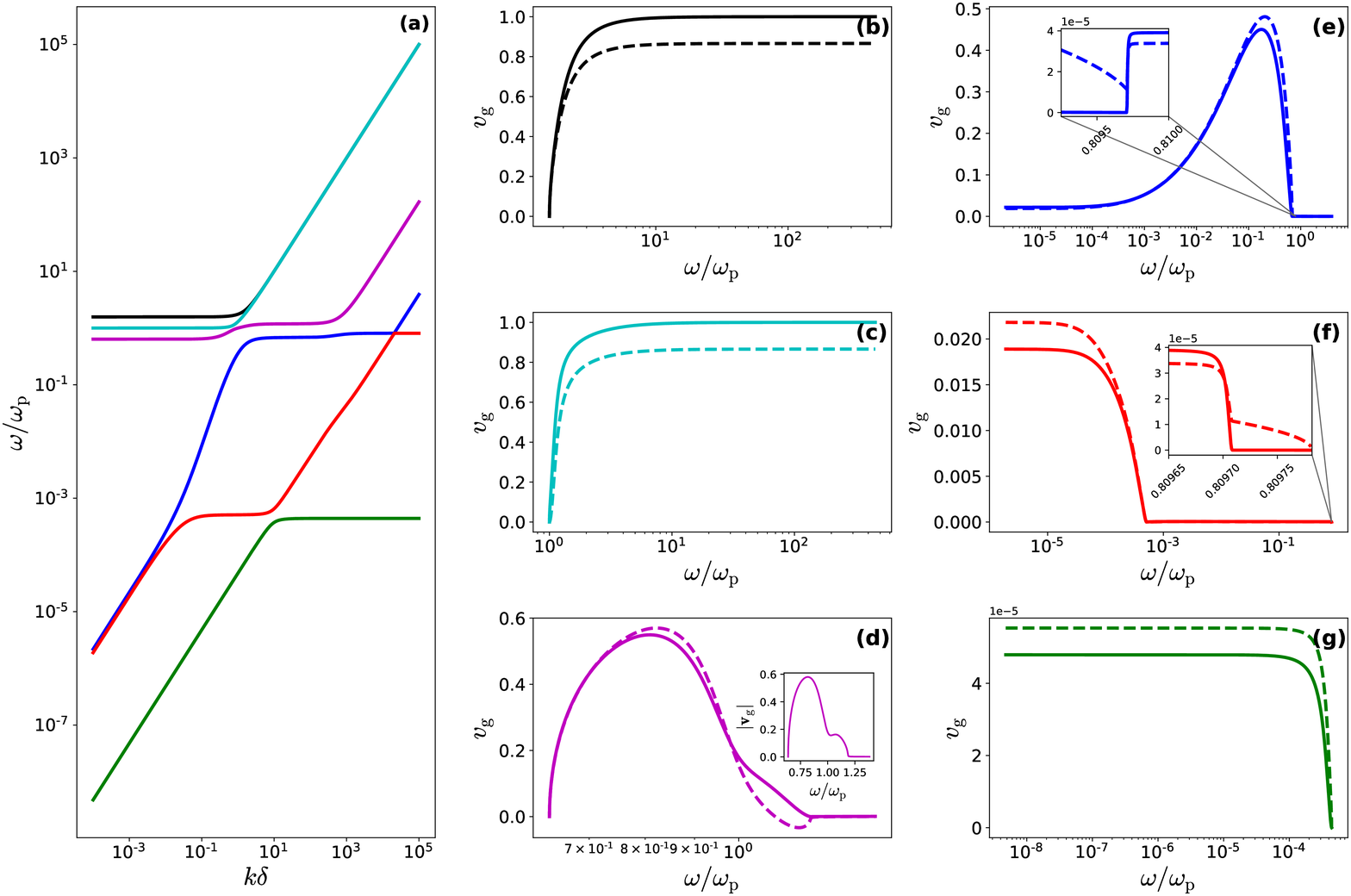}
\caption{(a) Dispersion diagram of Earth's magnetosphere conditions at an angle $\theta = \pi/6$ ($E < 1$, $c_\mathrm{s} < c_\mathrm{a}$). (b-g) Mode group speed components along the wavevector $\mathbf{v}_{\mathrm{g}} \cdot \hat{\mathbf{n}}$ (solid) and the magnetic field $\mathbf{v}_{\text{g}} \cdot \hat{\mathbf{b}}$ (dashed) for the (b) X mode, (c) O mode, (d) M mode, (e) F mode, (f) A mode, and (g) S mode, for varying frequency ranges based on cutoffs and/or resonances. The inset of (d) shows the magnitude of the M mode's group speed $|\mathbf{v}_{\mathrm{g}}|$, showing two peaks. The insets of (e) and (f) show the group speed behaviour near the AF avoided crossing.}
\label{fig:whistler-oblique}
\end{figure}

In Fig. \ref{fig:whistler-oblique}, the dispersion curve and the group speed of each mode are shown for Earth's magnetosphere parameters at a propagation angle $\theta = \pi/6$. Strong increases correspond to descending frequency whistling behaviour (abbreviated DFW), like the classical whistlers, and strong decreases to ascending frequency whistling behaviour (abbreviated AFW), like the ascending frequency and ion cyclotron whistlers.

In Fig. \ref{fig:whistler-oblique}(e), the F mode, which featured both the classical whistler and the ascending frequency whistler at parallel propagation, shows similar whistling behaviour, where a classical whistler for small frequencies is followed by an ascending frequency whistler. Near the electron cyclotron resonance however, the avoided crossing occurs and the group speed ``jumps" to a constant in the short wavelength limit where $\bar{\omega}^2 \simeq \bar{k}^2 w^2$ for the F mode, as can be seen in the inset. In (f), one can see that the A mode, which featured the ion cyclotron whistler at parallel propagation, still has this ascending frequency whistling behaviour near the ion cyclotron resonance, after which it vanishes at the electron cyclotron frequency, which is shown in the inset. Finally, in (g), one can see that the S mode now features ion cyclotron whistling behaviour at the ion cyclotron frequency due to the SA avoided crossing.

\subsubsection{Classical whistlers}
From Fig. \ref{fig:whistler-oblique}(e), it is clear that the F mode also describes classical whistler behaviour at oblique propagation angles in the frequency interval between the two resonance frequencies $\lambda I$ and $\lambda E$, at least in the regime $E < 1$, $c_\mathrm{s} < c_\mathrm{a}$. Like the parallel case, we can use a similar approximation in the general, sixth order dispersion relation to describe this whistling behaviour and expand textbook approximations to oblique angles. However, now we have to keep up to $\mathcal{O}(\bar{\omega}^6)$ in the dispersion relation to include the F mode. Once again, the ion contributions are limited to first order $\mathcal{O}(I)$. Additionally, at oblique angles the thermal speeds $v$ and $w$ also enter the equation. However, we will discard these thermal contributions for Earth's magnetosphere conditions, as $v$ is of similar order of magnitude as $I$ (and $v > w$) and the electron and ion sound speeds only appear squared. In fact, discarding these terms removes the S mode and thus tremendously simplifies the expression. It reduces the approximation to a first order equation in $\bar{\omega}^2$,
\begin{equation}\label{eq:obl-classical-freq}
\bar{\omega}^2 = \frac{\bar{k}^2 \left[ EI(1+\lambda^2) + \bar{k}^2 \left( EI + \lambda^2 E(E-I) \right) \right]}{1+2EI+\bar{k}^2 \left[ 2(1+2EI) + (1+\lambda^2) E(E-I) \right] + \bar{k}^4 (1+E^2)}.
\end{equation}
Whilst this approximation is reasonably good in the whistler region, it should be noted that it does not capture the avoided crossing behaviour at all \cite{DeJongheKeppens2020}. Nevertheless, starting from this expression, the phase and group speed expressions can be derived once again,
\begin{equation}
\mathbf{v}_{\mathrm{ph}} = \left[ \frac{EI(1+\lambda^2) + \bar{k}^2 \left( EI + \lambda^2 E(E-I) \right)}{1+2EI+\bar{k}^2 \left[ 2(1+2EI) + (1+\lambda^2) E(E-I) \right] + \bar{k}^4 (1+E^2)} \right]^{1/2} \hat{\mathbf{n}}
\end{equation}
and
\begin{equation}\label{eq:cw-gs-oblique}
\frac{\partial\bar{\omega}}{\partial\bar{\mathbf{k}}} = -\frac{\bar{k}}{\bar{\omega} P_\omega} \left[ \left( P_k - \frac{\lambda^2 P_\lambda}{\bar{k}^2} \right) \hat{\mathbf{n}} + \frac{\lambda P_\lambda}{\bar{k}^2}\ \hat{\mathbf{b}} \right]
\end{equation}
where
\begin{align}
P_\omega &= 1+2EI + \bar{k}^2 \left[ 2(1+2EI) + (1+\lambda^2)E(E-I) \right] + \bar{k}^4 (1+E^2), \label{eq:cw-po}\\
P_k &= 2\bar{k}^2 \left[ \bar{\omega}^2 (1+E^2) - E \left( I+\lambda^2 (E-I) \right) \right] \nonumber \\
&\quad + \bar{\omega}^2 \left[ 2(1+2EI) + (1+\lambda^2) E (E-I) \right] - (1+\lambda^2) EI, \label{eq:cw-pk}\\
P_\lambda &= \bar{k}^2 E \left[ \bar{\omega}^2 (E-I) - I - \bar{k}^2 (E-I) \right]. \label{eq:cw-pl}
\end{align}
Using eq. (\ref{eq:obl-classical-freq}) we can once again obtain a single-variable group speed expression. These expressions are already quite involved for such a simple approximation. Keeping any correction from thermal contributions complicates it even more. As it turns out, the first order ion correction is not as significant as in the parallel case discussed in Sec. \ref{sec:df-whistlers}. Therefore, the above expression can be reduced further by substituting $I = 0$. However, it seems that the ion correction becomes more important the further we deviate from parallel propagation.

\begin{figure}[!htb]
\includegraphics[width=\textwidth]{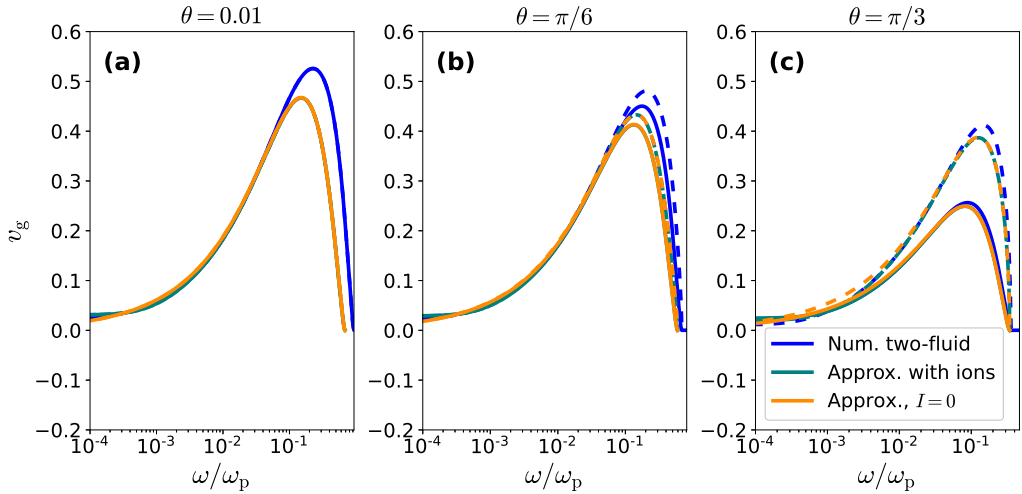}
\caption{Comparison of the oblique whistler group speed approximation (\ref{eq:cw-gs-oblique}) to (\ref{eq:cw-pl}) and (\ref{eq:obl-classical-freq}) and the exact two-fluid solution under magnetosphere conditions for different propagation angles: (a) $\theta = 0.01$, (b) $\theta = \pi/6$, and (c) $\theta = \pi/3$ ($\mathbf{v}_{\mathrm{g}} \cdot \hat{\mathbf{n}}-$solid, $\mathbf{v}_{\mathrm{g}} \cdot \hat{\mathbf{b}}-$dashed).}
\label{fig:ms-oblique-whistler}
\end{figure}

In Fig. \ref{fig:ms-oblique-whistler}, the evaluation of the approximation given by eqs. (\ref{eq:obl-classical-freq}) and (\ref{eq:cw-gs-oblique}) to (\ref{eq:cw-pl}) is shown for different angles alongside a numerical evaluation of the full dispersion relation group speed. As can be seen from the figure, the approximation is quite good in the classical whistler region, but underestimates the height of the peak.

\subsubsection{Ascending frequency whistlers}
\begin{figure}[!htb]
\includegraphics[width=\textwidth]{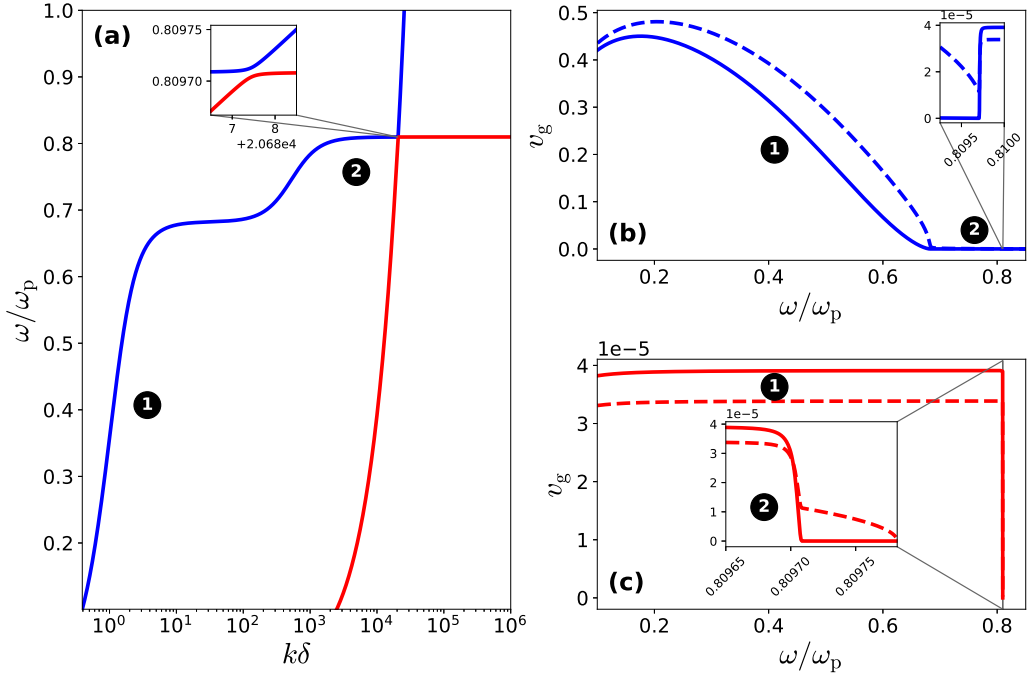}
\caption{(a) Dispersion diagram of the A and F modes in the ascending frequency whistler region for Earth's magnetosphere conditions at $\theta = \pi/6$. This is a zoom of Fig. \ref{fig:whistler-oblique}(a). The inset shows the avoided crossing. (b) F mode group speed. (c) A mode group speed. ($\mathbf{v}_{\mathrm{g}} \cdot \hat{\mathbf{n}}-$solid, $\mathbf{v}_{\mathrm{g}} \cdot \hat{\mathbf{b}}-$dashed)}
\label{fig:afw-oblique}
\end{figure}

At parallel propagation, the ascending frequency whistler occurs in the F mode when the frequency approaches the electron cyclotron resonance $E$ asymptotically. For the magnetosphere regime ($E < 1$, $c_\mathrm{s} < c_\mathrm{a}$), the S mode crosses the F mode at a frequency close to $E$. At oblique angles, this means that the S, A, and F modes are affected, and that there is an avoided crossing in the frequency interval where we expect the ascending frequency whistler to be. Fig. \ref{fig:afw-oblique}(a) shows a closer look at the AFW region where the avoided crossing between the A and F modes occurs. Now note that the F mode has two plateau-like regions. This is due to the fact that at oblique angles the resonances of the cold and warm plasma differ. The highest plateau occurs near $\bar{\omega} = \lambda E$, where we expect the avoided crossing to be, and the associated whistling behaviour. This is a warm plasma resonance. An approximate value for the lowest plateau on the other hand can be found by calculating the resonance limit for a cold plasma, and neglecting ion contributions, to get \cite{Keppens2019_coldei}
\begin{equation}\label{eq:freq-glh-approx}
\bar{\omega}^2 \simeq \frac{1}{2} \left( 1+E^2 - \sqrt{(1+E^2)^2 - 4\lambda^2 E^2} \right).
\end{equation}
This frequency lies between zero (for $\lambda = 0$) and the parallel resonance limit $\bar{\omega}^2 = E^2$ (for $\lambda = 1$). Near perpendicular propagation the ion contributions should not be neglected though, and we find a lower bound of
\begin{equation}\label{eq:freq-lh}
\bar{\omega}_{\mathrm{LH}}^2 = \frac{1}{2} \left( 1+E^2+I^2 - \sqrt{(1+E^2+I^2)^2 - 4EI(1+EI)} \right),
\end{equation}
which is commonly referred to as the lower hybrid frequency $\bar{\omega}_{\mathrm{LH}}$ \cite{Keppens2019_coldei} and is $\omega_{\mathrm{LH}} = 90\ \mathrm{kHz}$ for the typical reference Earth's magnetosphere parameters. Looking at Fig. \ref{fig:afw-oblique}(b), the lowest plateau marks the end of the large scale AFW behaviour. Hence, this strong AFW behaviour occurs near the cold plasma resonance that falls from the electron cyclotron frequency at parallel propagation to the lower hybrid frequency (\ref{eq:freq-lh}) at perpendicular propagation, and is approximately given by (\ref{eq:freq-glh-approx}), except near $\lambda = 0$. This behaviour is described adequately by the cold plasma limit.

Above the cold plasma resonance given approximately by eq. (\ref{eq:freq-glh-approx}), the insets of Fig. \ref{fig:afw-oblique}(b) and (c) show that near the electron cyclotron frequency (the F mode's highest plateau) there is also AFW behaviour in the F mode, albeit on a much smaller scale, and the A mode, due to the avoided crossing. In the F mode, the avoided crossing causes small scale DFW behaviour immediately after this small scale AFW behaviour.

If one would like to analytically approximate the small scale whistling behaviour near the resonance (region 2 in Fig. \ref{fig:afw-oblique}), a short wavelength approximation similar to the one used in the parallel case (see Sec. \ref{sec:afw-pl}) can be obtained for oblique angles by keeping only terms of order $\mathcal{O}(\bar{k}^8)$ and $\mathcal{O}(\bar{k}^6)$ in the general two-fluid dispersion relation \cite{GoedbloedKeppensPoedts2019}. Assuming $\lambda I < \bar{\omega} < \lambda E$ and applying the approximation $\bar{\omega} = \lambda E$ in all but the vanishing factor, an extension of the previous phase and group speed expressions to oblique angles is obtained, including ion and thermal contributions,
\begin{equation}
\mathbf{v}_{\mathrm{ph}} = \lambda v \sqrt{\frac{2E(E+I)}{2v^2+E(E+I)(1-\lambda^2)}} \left( 1-\frac{\bar{\omega}}{\lambda E} \right)^{1/2}\ \hat{\mathbf{n}}
\end{equation}
and
\begin{equation}\label{eq:afwgroup-simpleobl}
\frac{\partial\bar{\omega}}{\partial\bar{\mathbf{k}}} = v \sqrt{\frac{2E(E+I)}{2v^2+E(E+I)(1-\lambda^2)}} \left( 1 - \frac{\bar{\omega}}{\lambda E} \right)^{1/2} \left\{ 2 \lambda \left( 1 - \frac{\bar{\omega}}{\lambda E} \right) \hat{\mathbf{n}} + \left( \hat{\mathbf{b}} - \lambda \hat{\mathbf{n}} \right) \right\}.
\end{equation}
These expressions reduce to eqs. (\ref{eq:afwphase-ioncorr}) and (\ref{eq:afwgroup-ioncorr}) at parallel propagation where $\lambda = \pm 1$ and $\hat{\mathbf{b}} = \lambda\hat{\mathbf{n}}$. Interesting to note is that the oblique behaviour is influenced by the electron sound speed $v$ whilst the ion sound speed $w$ drops out of the expression.

\begin{figure}[!htb]
\includegraphics[width=\textwidth]{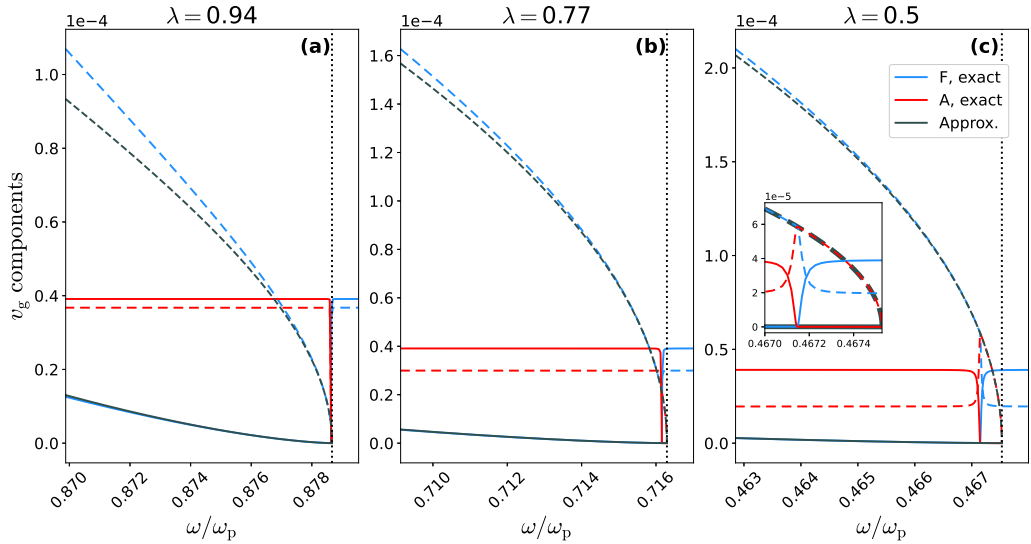}
\caption{The A and F mode group speeds for Earth's magnetosphere conditions are shown in red and blue, respectively, and the AFW approximation in black for (a) $\lambda = 0.94$, (b) $\lambda = 0.77$, and (c) $\lambda = 0.5$, in the interval $(0.99\ \lambda E, 1.001\ \lambda E)$. ($\mathbf{v}_{\mathrm{g}} \cdot \hat{\mathbf{n}}-$solid, $\mathbf{v}_{\mathrm{g}} \cdot \hat{\mathbf{b}}-$dashed) The approximation is unaware of the avoided crossing and follows the F mode before and the A mode after the avoided crossing. A close-up of this behaviour is shown in the inset of (c). The resonance $\lambda E$ is indicated by the vertical dotted black line.}
\label{fig:whistler-lamdep}
\end{figure}

In Fig. \ref{fig:whistler-lamdep}, the group speed approximation (\ref{eq:afwgroup-simpleobl}) is shown alongside the exact solutions of the general dispersion relation for the A and F mode. The approximation is unaware of the avoided crossing and thus approximates the F mode up until the avoided crossing, whereafter it approximates the A mode closer to the resonance. This can be seen most clearly in the inset of panel (c). The approximation is reasonably good near the resonance, but becomes progressively worse for smaller values of $\bar{\omega}$, where the wavenumber is also smaller and the large wavenumber approximation breaks down.

\subsubsection{Ion cyclotron whistlers}
\begin{figure}[!htb]
\includegraphics[width=\textwidth]{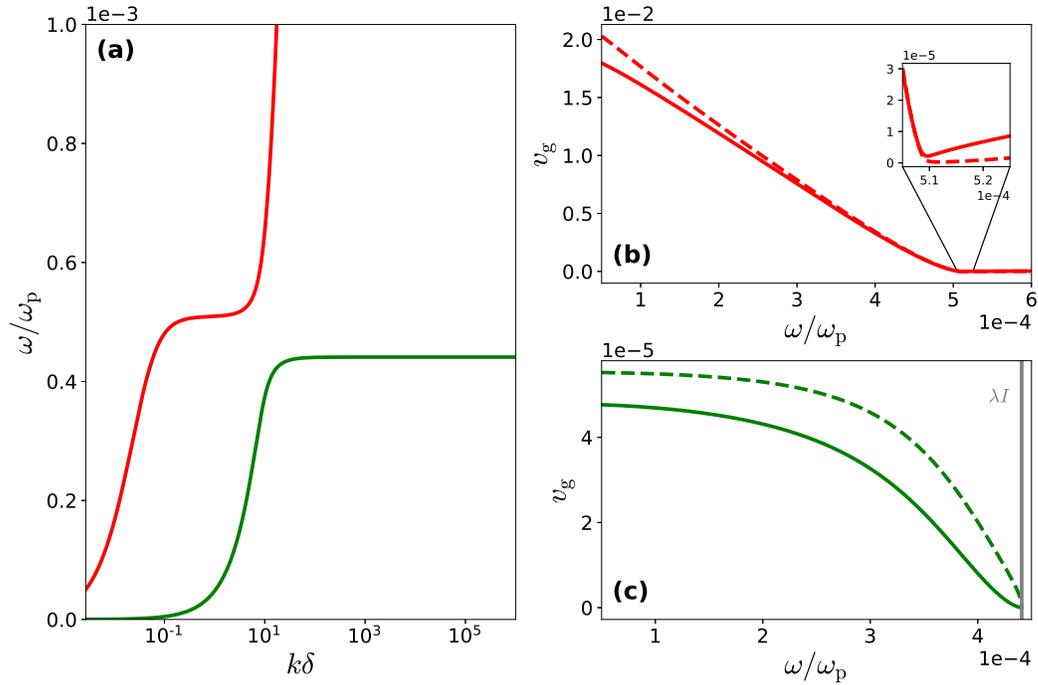}
\caption{(a) Dispersion diagram of the S and A modes in the ion cyclotron whistler region for Earth's magnetosphere conditions at $\theta = \pi/6$. This is a zoom of Fig. \ref{fig:whistler-oblique}(a). (b) A mode group speed. (c) S mode group speed. ($\mathbf{v}_{\mathrm{g}} \cdot \hat{\mathbf{n}}-$solid, $\mathbf{v}_{\mathrm{g}} \cdot \hat{\mathbf{b}}-$dashed) Note that the vertical axis in all panels has a multiplicative scale factor indicated at their top left corner.}
\label{fig:icw-oblique}
\end{figure}

The ion cyclotron resonance is given at any angle by $\bar{\omega} = \lambda I$. Since the ion cyclotron whistler occurs at frequencies near $I$ at parallel propagation, we look for similar behaviour near $\lambda I$ at oblique angles. Once again, at parallel propagation two modes cross near $\bar{\omega} = I$ in the magnetosphere regime ($E < 1$, $c_\mathrm{s} < c_\mathrm{a}$), namely the S and A modes. Just like the previous case, an avoided crossing appears at a small deviation from parallel propagation. This avoided crossing can still be seen at a larger propagation angle in Fig. \ref{fig:icw-oblique}(a), although the modes no longer approach each other extremely closely.

When moving further away from parallel propagation, it becomes clear that there are two frequencies of interest, $\bar{\omega} = I$ and $\bar{\omega} = \lambda I$. This is illustrated in Fig. \ref{fig:icw-oblique} for an angle $\theta = \pi/6$. Here, you can see in (a) that the S mode approaches $\bar{\omega} = \lambda I$ asymptotically and the A mode has a plateau near $\bar{\omega} = I$. The exact value of the plateau is given by the lowest resonance of the cold plasma, which is very close to the ion cyclotron frequency at all angles except near perpendicular propagation, where the A mode vanishes. The S mode's group speed is visualised near the resonance in (c), where it varies relatively rapidly. There is indeed AFW behaviour, although the group speed is quite small. Hence, the avoided crossing leads to the occurrence of the oblique ion cyclotron whistler in the S mode near $\bar{\omega} = \lambda I$. The group speed of the A mode near the plateau is shown in the inset of (b). It appears to have both AFW and minor DFW behaviour near this cold plasma resonance.

\subsubsection{High-frequency whistlers}\label{sec:oblique-hfw}
For Earth's magnetosphere, Fig. \ref{fig:whistler-oblique}(b) shows that the X mode's oblique behaviour is largely unaltered from parallel propagation, in Fig. \ref{fig:hfw}(b). The M and O modes, which cross in the O mode's whistling region at parallel propagation, switch their cutoffs such that $\omega_\mathrm{M} < \omega_\mathrm{O}$ at all wavenumbers. Since the cutoffs are where the whistling behaviour occurs, this behaviour is heavily influenced. This is especially pronounced in the oblique M mode, which can be seen by comparing Fig. \ref{fig:hfw}(c,d) and Fig. \ref{fig:whistler-oblique}(c,d). In the parallel case, both the M and O mode only feature DFW behaviour whereas at oblique angles the M mode shows strong DFW and AFW behaviour. Both behaviours occur in the frequency range between the lower cutoff frequency
\begin{equation}\label{eq:lowercutoff}
\omega_\mathrm{l} = \left[ 1 + \frac{1}{2} (E^2+I^2) - \frac{1}{2} |E-I| \sqrt{(E+I)^2+4} \right]^{1/2} \omega_\mathrm{p}
\end{equation}
and the highest cold plasma resonance given approximately by \cite{Keppens2019_coldei}
\begin{equation}\label{eq:freq-guh-approx}
\bar{\omega}^2 \simeq \frac{1}{2} \left( 1+E^2 + \sqrt{(1+E^2)^2 - 4\lambda^2 E^2} \right),
\end{equation}
ignoring ion contributions. This value rises from the plasma frequency at parallel propagation ($\lambda = 1$) to the upper hybrid frequency $\bar{\omega}_{\mathrm{UH}}$, given by \cite{Keppens2019_coldei}
\begin{equation}\label{eq:freq-uh}
\bar{\omega}_{\mathrm{UH}}^2 = \frac{1}{2} \left( 1+E^2+I^2 + \sqrt{(1+E^2+I^2)^2 - 4EI(1+EI)} \right),
\end{equation}
at perpendicular propagation ($\lambda = 0$). As shown in Fig. \ref{fig:whistler-oblique}(d), the magnitude of the group speed actually has two peaks, which are separated by the plasma frequency
($\omega_\mathrm{p} = 5.64\ \mathrm{MHz}$, $\omega_\mathrm{l} = 3.59\ \mathrm{MHz}$, $\omega_{\mathrm{UH}} = 7.72\ \mathrm{MHz}$ for the typical reference Earth's magnetosphere parameters). The O mode's whistling behaviour is situated directly above its cutoff at the plasma frequency and is limited to DFW behaviour.

\subsubsection{Pair plasma whistlers}\label{sec:oblique-pair}
\begin{figure}[!htb]
\includegraphics[width=\textwidth]{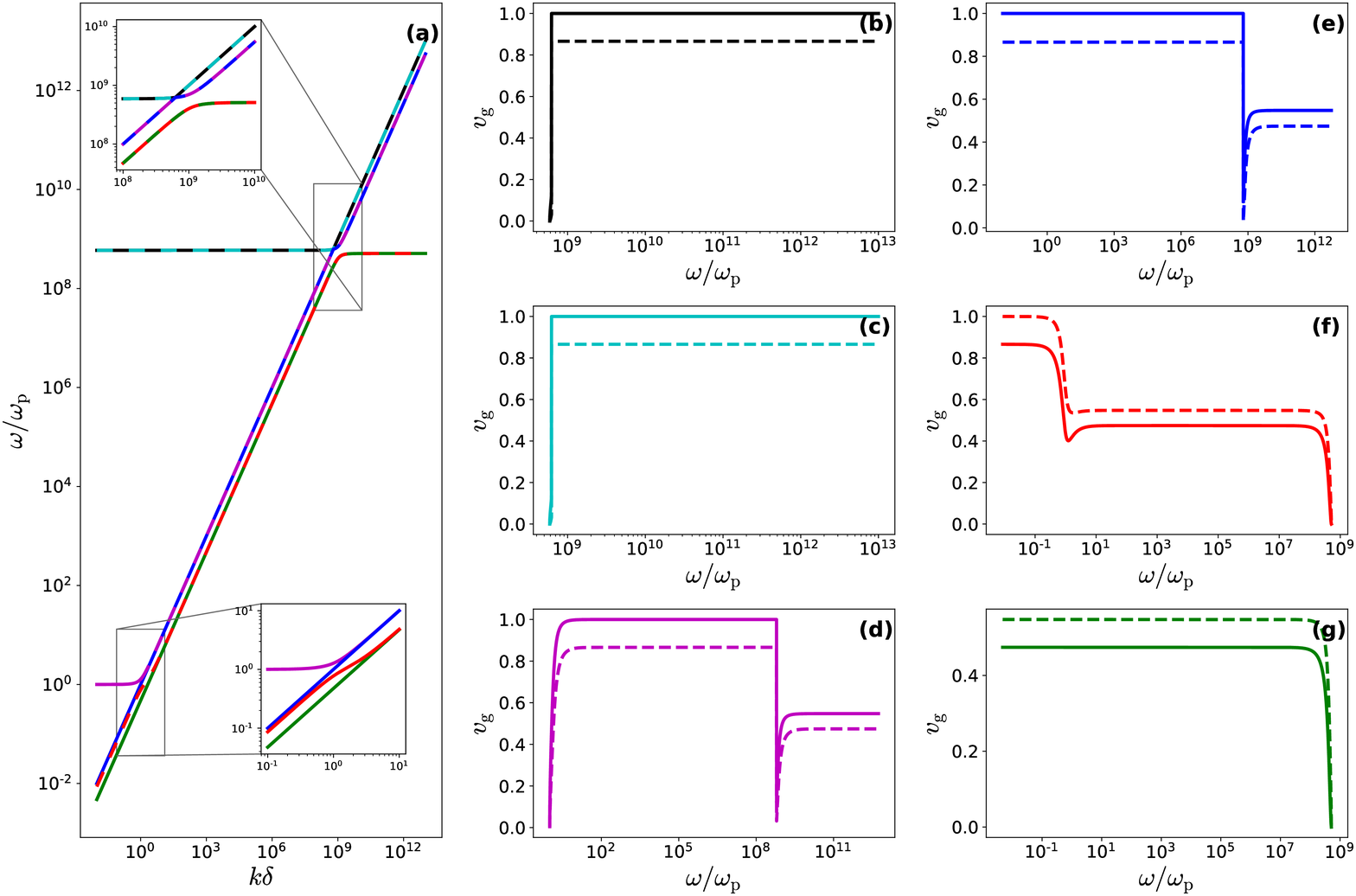}
\caption{(a) Dispersion diagram of a pair plasma under pulsar magnetosphere conditions at an angle $\theta = \pi/6$ ($E \simeq 5.89\times 10^{8}$). Note that due to the extreme conditions inherent in the pulsar magnetosphere, modes that virtually overlap are shown as lines that show both associated colours, even in some of the zoomed inset views. (b-g) Mode group speed components along the wavevector $\mathbf{v}_{\mathrm{g}} \cdot \hat{\mathbf{n}}$ (solid) and the magnetic field $\mathbf{v}_{\text{g}} \cdot \hat{\mathbf{b}}$ (dashed) for the (b) X mode, (c) O mode, (d) M mode, (e) F mode, (f) A mode, and (g) S mode, for varying frequency ranges based on cutoffs and/or resonances.}
\label{fig:pair-oblique}
\end{figure}

Returning to a pulsar magnetosphere environment for a moment, it was shown that even though a pair plasma does not feature classical whistlers, it does have AFW behaviour in the A and F modes at parallel propagation. At oblique angles, the frequency ordering in a pair plasma \cite{Keppens2019_warmpair} ensures that it are now the S and A modes that both approach the cyclotron resonance $\lambda E$. Fig. \ref{fig:pair-oblique}, (f) and (g), shows that both the S and A mode display AFW behaviour near the resonance. Additionally, the A mode also features AFW behaviour near the plasma frequency ($\bar{\omega} = 1$), followed by smaller scale DFW behaviour. The same behaviour, AFW followed by DFW, is also present in the F and M modes, which nearly coincide for short wavelengths, near the cyclotron frequency ($\bar{\omega} = E$). Furthermore, the O and X modes also feature DFW behaviour near their cutoff frequencies. Note that since the plasma frequency is well above the audible frequency range ($\omega_\mathrm{p} \simeq 29.8\ \mathrm{GHz}$), none of these behaviours actually occurs in the audible range. Hence, whilst a pair plasma does not feature classical whistlers at parallel propagation, there is still a lot of interesting whistling behaviour occuring across all modes.

\subsection{Cross-field whistlers}
\begin{figure}[!htb]
\includegraphics[width=\textwidth]{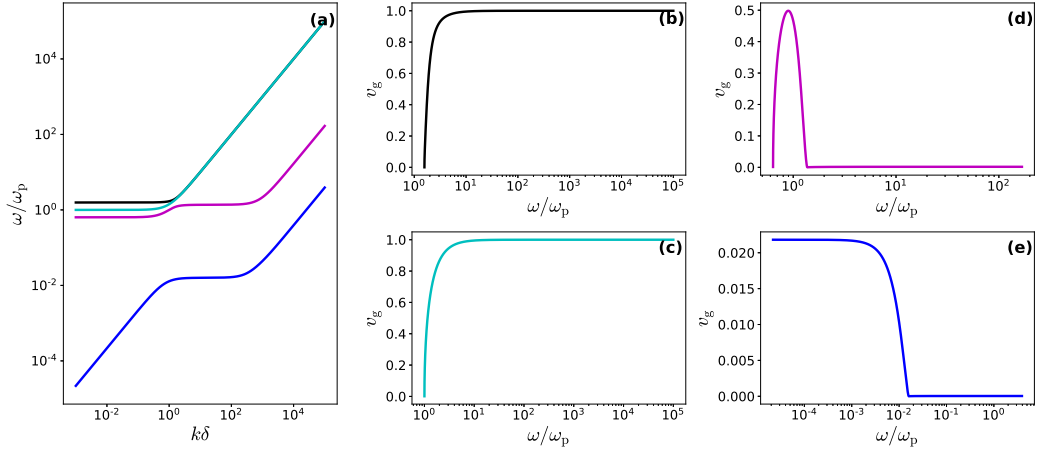}
\caption{(a) Dispersion diagram of the F, M, O, and X modes for Earth's magnetosphere at perpendicular propagation. (b-e) Mode group speeds along the wavevector, perpendicular to the magnetic field, for the (b) X mode, (c) O mode, (d) M mode, and (e) F mode, for varying frequency ranges based on cutoffs.}
\label{fig:groupspeed-perp}
\end{figure}

In the limit of perpendicular propagation, the S and A modes no longer propagate, in exact correspondence with the known MHD property of slow and Alfv\'en waves, resulting in a different factorisation of the dispersion relation \cite{DeJongheKeppens2020}. Additionally, there are no resonances at perpendicular propagation, where AFW behaviour occurred at parallel and oblique angles. However, computing the group speed of the F mode numerically, which is then perpendicular to the magnetic field, shows that small scale AFW behaviour occurs at the previously introduced lower hybrid frequency (\ref{eq:freq-lh}). The group speed behaviour of all four perpendicularly propagating modes is shown for Earth's magnetosphere conditions in Fig. \ref{fig:groupspeed-perp}. In particular, the F mode's steep decrease in group speed, shown in (e), is caused by the presence of a plateau in the F mode dispersion curve at the lower hybrid frequency, shown in (a). Note that the group speed difference is of the same scale as the parallel ion cyclotron whistler in Fig. \ref{fig:ioncycl-whistler}. However, this AFW whistling occurs near the lower hybrid frequency at $90\ \mathrm{kHz}$, which is outside of the audible frequency range. The high-frequency M, O, and X modes again show DFW behaviour near their cutoffs, similar to their oblique behaviour. For the M mode, this DFW behaviour is followed by equally strong AFW behaviour as the frequency approaches the upper hybrid frequency $\bar{\omega}_{\mathrm{UH}}$. Unlike in the oblique case, the whistling behaviour of the M mode does not show any changes at the plasma frequency.

\section{Warm extension of the Appleton-Hartree relation}
In most whistler applications discussed so far, the pairs involved from the six mode dispersion relation are the S, A, and F low-frequency ones. As a different application of the general, sixth order dispersion relation, we can also study high-frequency waves. In magneto-ionic theory this is usually done using the Appleton-Hartree relation, which is a cold approximation. Relatively recently, \citeA{Bawaaneh2013} derived a warm version of the Appleton-Hartree equation as a dispersion relation combining the frequency and the index of refraction. To obtain the warm Appleton-Hartree equation in a manner consistent with the S, A, F, M, O, X labelling scheme, it suffices to consider the $\mu = 0$ limit of the general two-fluid dispersion relation in $\bar{\omega}$ and $\bar{k}$. For the cold ion-electron case, where only 5 mode pairs exist, this was already done in \citeA{Keppens2019_coldei}, where the equation was shown to retain 4 out of the 5 mode pairs. The $\mu = 0$ limit implies that $I = 0$ and $c_\mathrm{s}^2 = w^2$. With these substitutions the fully general dispersion relation factors out a trivial mode $\bar{\omega}^2 = 0$ just like the cold case and it separates the ion and electron sound speeds $v$ and $w$ resulting in the dispersion relation
\begin{equation} \label{eq:general-ah-omegak}
\begin{aligned}
0 = &\bar{\omega}^2 (\bar{\omega}^2 - \bar{k}^2 w^2) \{ \bar{\omega}^8 - \bar{\omega}^6 \left[ 3 +E^2 +\bar{k}^2(2+v^2) \right] \\ &+ \bar{\omega}^4 \left[ 3 +E^2 +\bar{k}^2 \left( 4 +2v^2 +2E^2 +\lambda^2 E^2 v^2 \right) +\bar{k}^4 (1+2v^2) \right] \\ &- \bar{\omega}^2 [ 1 +\bar{k}^2 \left( 2 +v^2 +(1+\lambda^2)E^2 \right) +\bar{k}^4 \left( 1 +2v^2 +E^2 +2\lambda^2 E^2 v^2 \right) \\ &+\bar{k}^6 v^2 ] + \bar{k}^4 \lambda^2 E^2 (1+\bar{k}^2v^2) \}.
\end{aligned}
\end{equation}
It can be argued that the factor $\bar{\omega}^2-\bar{k}^2 w^2$ should be disregarded or replaced by $\bar{\omega}^2$ because the $\mu = 0$ limit implies that the ions are infinitely heavy and therefore immobile. Hence, their thermal speed $w$ should also be $w = 0$. Mathematically, it is interesting that it factors out anyway. This leaves the polynomial of fourth degree in $\bar{\omega}^2$ as a generalisation of the Appleton-Hartree dispersion relation to the warm case. Expression ($24$) in \citeA{Bawaaneh2013}, ignoring collision terms, can be reordered to retrieve eq. (\ref{eq:general-ah-omegak}).

It is easily checked that substituting $v = 0$, i.e. setting the electron thermal speed to zero, in this polynomial retrieves eq. (40) from \citeA{Keppens2019_coldei}, which they showed to be equivalent to the Appleton-Hartree relation, usually written as a function of the refractive index $n = ck/\omega = \bar{k}/\bar{\omega}$. This usual way of expressing this relation is only for historical reasons preferred and is possible because the polynomial is of second degree in $\bar{k}^2$ in the cold case. However, including a non-zero thermal velocity $v$ adds two terms of third degree in $\bar{k}^2$. Writing $n^2$ as an explicit function of the other variables is thus more involved. Additionally, as pointed out in \citeA{Keppens2019_coldei}, this hopelessly confuses the unique and unambiguous S, A, F, M, O, X wave labelling and identification.

Nevertheless, adopting the usual Appleton-Hartree variables $X$ and $Y$ defined as $X = 1/\bar{\omega}^2$ and $Y = E/\bar{\omega}$, and introducing a new variable $V = v/\bar{\omega}$ related to the thermal velocity, the warm equivalent of the Appleton-Hartree relation can be expressed as a third degree polynomial in $n^2$
\begin{equation}\label{eq:general-appletonhartree}
\begin{aligned}
0 = &n^6 V^2 (1-Y^2 \cos^2\theta) - n^4 [ X + 2V^2 (1-X) - X^2 (1-Y^2\cos^2\theta) \\ &- Y^2 (X + 2V^2\cos^2\theta) ] + n^2 [ 2X (1-V^2) + V^2 (1-Y^2\cos^2\theta) + 2X^2 (X-2) \\ &- 2XY^2 + X^2V^2 + (1+\cos^2\theta) X^2Y^2 ] - X (1-X) \left[ (1-X)^2 - Y^2 \right].
\end{aligned}
\end{equation}
The formula for roots of a cubic polynomial can then be used to write the squared refractive index $n^2$ as a function of $X$, $Y$, $V$, and $\theta$. However, the expression does not simplify significantly. Thus, it is omitted here in favor of the polynomial expression. It is to be noted that roots of arbitrary degree polynomials are routinely computed numerically anyway.

Similar to how the full dispersion relation factorised for parallel and perpendicular propagation, the generalised Appleton-Hartree relation (\ref{eq:general-appletonhartree}) factorises if we substitute $\theta = 0$ or $\theta = \pi/2$. For parallel propagation ($\theta = 0$), the expression becomes
\begin{equation}\label{eq:warm-ah-pl}
0 = [n^2 V^2 - X(1-X)] [n^2 (1-Y) - (1-X-Y)] [n^2 (1+Y) - (1-X+Y)].
\end{equation}
Comparing this to the parallel factorisation in \citeA{DeJongheKeppens2020}, the last two factors come from the quartic branch whilst the first factor comes from the quadratic branch. The discarded factor $\bar{\omega}^2 - \bar{k}^2 w^2$ also came from the quadratic branch. Note that the substitution $\omega \rightarrow -\omega$ transforms the variables as $X \rightarrow X$, $Y \rightarrow -Y$ and $V \rightarrow -V$. Therefore, the last two factors in this relation mix forward-backward wave types as was also pointed out in \citeA{Keppens2019_coldei} and \citeA{DeJongheKeppens2020}. This intrinsic mixing of forward-backward wave types seems to be a recurring habit in all the plasma physics literature, and it is ``justified" from the fact that at these special parallel or perpendicular orientations, some wave modes become then classifiable as left or right hand polarised wavetypes. It is to be noted that the forward-backward pairing, which we from now on advocate as preferential, carries over to moving reference frames, and has been crucial in rigorously analysing MHD waves in stationary configurations with spatially varying equilibrium conditions \cite{GoedbloedKeppensPoedts2019}.

Analogously, for perpendicular propagation ($\theta = \pi/2$) the factorisation is
\begin{equation}
\begin{aligned}
0 = (n^2-&1+X) \\
&\times \left\{ n^4 V^2 - n^2 \left[ V^2 (1-X) + X (1-X-Y^2) \right] + X \left[ (1-X)^2-Y^2 \right] \right\}.
\end{aligned}
\end{equation}
Once again, this result corresponds to the perpendicular factorisation in \citeA{DeJongheKeppens2020}. The linear branch matches without simplification whilst the cubic branch reduces to the quadratic expression in $n^2$ here and the factor $\bar{\omega}^2 - \bar{k}^2 w^2$.

\begin{figure}[!htb]
\centering
\includegraphics[width=\textwidth]{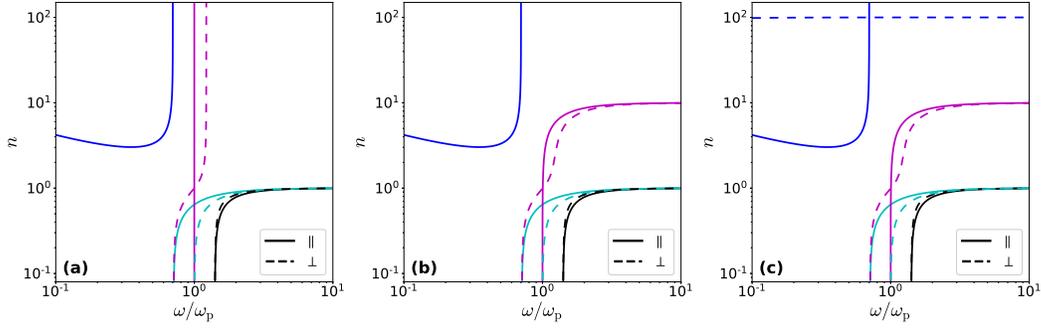}
\caption{(a) The (cold) Appleton-Hartree relation in an $n(\bar{\omega})$-diagram. (b) The extended warm Appleton-Hartree equation (\ref{eq:general-appletonhartree}) with $v = 0.1$. (c) The corresponding modes from the full dispersion diagram for an ion-electron plasma ($\mu \simeq 1/1836$) with $v = 0.1$ and $w = 0.01$. All cases use $E = 0.7$ and are shown in the limits of parallel (solid) and perpendicular (dashed) propagation.}
\label{fig:ahl-comparison}
\end{figure}

Finally, a closer look at how the warm version improves upon the cold Appleton-Hartree equation is visualised in Fig. \ref{fig:ahl-comparison}. Here, (a) shows the cold Appleton-Hartree relation, (b) shows the warm extension, and (c) shows an evaluation of the F, M, O, and X modes in the full dispersion relation for reference. For the two electromagnetic O and X modes, and the F mode, there does not seem to be a noticable improvement from cold to warm. In fact, the short wavelength (large $\bar{k}$) limit of the F mode is not captured correctly for oblique and perpendicular propagation. It should be $n^2 \simeq 1/w^2$, but $w$ only appears in a different factor in eq. (\ref{eq:general-ah-omegak}). However, the M mode (purple), related to the Langmuir wave, now exhibits a more correct high-frequency behaviour. Comparing Fig. \ref{fig:ahl-comparison}(b) and (c), all four modes seem to follow the correct behaviour now, except for the F mode at oblique and perpendicular propagation.

Fig. \ref{fig:ahl-comparison} only shows the parallel and perpendicular case, but it was pointed out in \citeA{DeJongheKeppens2020} that all crossings between modes become avoided crossings at oblique angles. This applies to the parallel MO crossing in Fig. \ref{fig:ahl-comparison} and also to any crossing appearing for higher values of $E$. If the MO crossing in (a) is replaced by an avoided crossing, the curves represent how the oblique M and O modes behave at near-parallel propagation before reconnecting and thus crossing at exactly parallel propagation. This recovers Fig. $4.37$ in \citeA{GurnettBhattacharjee2005}.

To conclude this section, we discuss the well-known Faraday rotation \cite{Bittencourt2004, GurnettBhattacharjee2005, ThorneBlandford2017, Keppens2019_coldei} in the cold and warm Appleton-Hartree equations. This effect occurs when we take a superposition of the electromagnetic O and X modes, with a frequency above the upper cutoff frequency $\omega_\mathrm{u}$, which is
\begin{equation}
\omega_\mathrm{u}^2 = 1 + \frac{E^2}{2} + \frac{E}{2} \sqrt{E^2+4}
\end{equation}
in both the cold and the warm Appleton-Hartree relation. Since the O and X modes have a different wavenumber for a given frequency, and thus a different phase speed, the resulting electric field of this wave will rotate as the wave propagates. This is usually discussed at parallel propagation, where the dispersion relation factorises as eq. (\ref{eq:warm-ah-pl}). In this equation, the last two factors correspond to the electromagnetic X and O modes, respectively.

\begin{figure}[!htb]
\centering
\includegraphics[width=\textwidth]{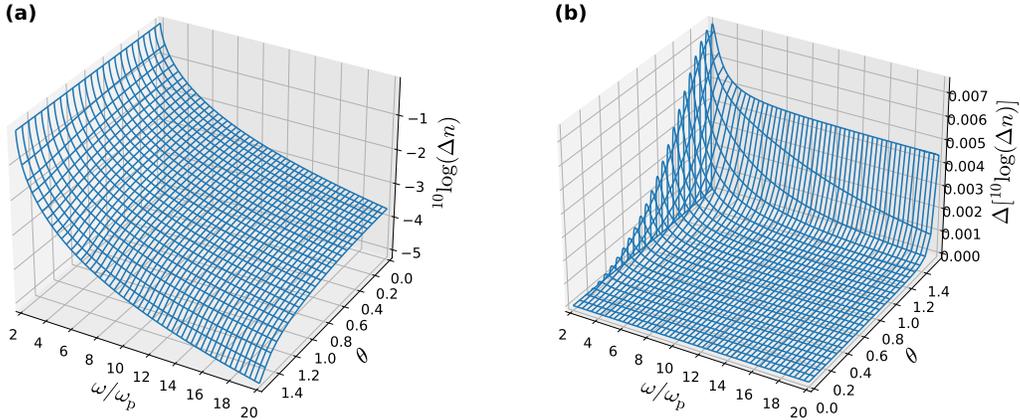}
\caption{(a) The Faraday rotation angle $\psi$ present in the warm Appleton-Hartree equation is quantified using $\psi \sim \Delta n = n_\mathrm{O} - n_\mathrm{X}$. (b) The influence of the warm extension to the Appleton-Hartree relation on the Faraday rotation angle $\psi$ is quantified as the difference of the logarithms of the rotation angles, $\log(n_\mathrm{O} - n_\mathrm{X})_\mathrm{warm} - \log(n_\mathrm{O} - n_\mathrm{X})_\mathrm{cold}$. For both panels parameter values were set to $E = 1.5$ and $v = 0.1$. Note that the $\theta$-axis is differently oriented for panels (a) and (b).}
\label{fig:faraday}
\end{figure}

The angular rotation rate $\psi$ can be quantified using the index of refraction for both modes, denoted $n_\mathrm{O}$ and $n_\mathrm{X}$, because $\psi$ is proportional to $\Delta n$, $\psi \sim n_\mathrm{O} - n_\mathrm{X}$. At parallel propagation, it gives identical results for the cold \cite{Keppens2019_coldei} and warm Appleton-Hartree relations, and is approximately
\begin{equation}
n_\mathrm{O} - n_\mathrm{X} \simeq \frac{E}{\bar{\omega}^3}.
\end{equation}
Numerically, it can be evaluated at any angle and any frequency. Doing so for the warm Appleton-Hartree expression, the result is shown in Fig. \ref{fig:faraday}(a). It looks almost identical to the general cold ion-electron case reported in \citeA{Keppens2019_coldei}. The difference of this numerical evaluation of the warm and the cold Appleton-Hartree relation is shown in Fig. \ref{fig:faraday}(b). As pointed out earlier, there is no difference at parallel propagation. The largest deviation between the cold and warm Appleton-Hartree expression occurs near perpendicular propagation. Even there, the difference is negligible though. Therefore, we conclude that the warm Appleton-Hartree relation derived above does not offer any significant advantage over its cold equivalent with respect to Faraday rotation.

\section{Conclusion}
Since whistler waves were originally observed as travelling parallel to the magnetic field, most plasma physics textbooks focus on approximating this behaviour at parallel propagation, discarding ion terms. However, more recent observations proved the existence of whistler waves at oblique angles. Using a two-fluid formalism, we showed that the parallel textbook approximations can be meaningfully extended to oblique angles. In the S, A, F, M, O, X labelling scheme, the ``classical" descending frequency whistler occurs in the F mode for Earth's magnetosphere conditions, both at parallel and oblique propagation. The ascending frequency and ion cyclotron whistlers, occurring in the F and A modes respectively at parallel propagation under Earth's magnetosphere conditions, are affected by the introduction of avoided crossings at oblique angles. The location of these avoided crossings in whistling regions effectively splits whistling behaviour across two modes at oblique angles. At perpendicular propagation, it turned out that the remaining F mode also features small scale ascending frequency whistling behaviour. Furthermore, the high-frequency M, O, and X modes also show whistler-like behaviour at all angles, albeit near their cutoffs, which lie outside of the audible frequency range.

Due to the avoided crossings, the whistling behaviour across all modes depends strongly on the parameters. Whilst the analysis in this paper is representative of Earth's magnetosphere, it will already be different for another magnetosphere in our solar system: Jupiter's. Modelling Jupiter as a magnetic dipole with a dipole moment of $1.584\times 10^{20}\ \mathrm{T}\ \mathrm{m}^3$ \cite{Milone2014}, neglecting its offset for simplicity, and using the ingress peak electron density from \citeA{Hinson1997} results in an electron cyclotron frequency of $E \simeq 5$ for a proton-electron plasma, which describes a different regime than Earth's magnetosphere ($E < 1$). Entirely different setups such as pair plasmas in pulsar magnetospheres are very distinct as well. In such a pair plasma there are no classical descending frequency whistlers at parallel propagation, but these plasmas do have modes with ascending frequency whistling behaviour at parallel propagation. At oblique angles, they even show a wide variety of whistling behaviour across all modes.

Additionally, the damping-free Appleton-Hartree equation was extended to incorporate the effect of a non-zero thermal electron velocity. Although this extension does not improve much upon the approximations of the F, O, and X modes, it does introduce an improved description of the M mode, related to the textbook Langmuir wave, capturing its unique high-frequency $\bar{\omega}^2 \simeq \bar{k}^2v^2$ behaviour. The polynomial form is more involved than for the cold case and consequently, explicit expressions of the refractive index, although possible, are not very insightful.

Whilst the employed two-fluid treatment is complete, all intricacies of damping effects \cite{Bell2002, Ma2017, Hsieh2018} are inherently absent, although collisional damping, which may be strong for whistler waves \cite{Crabtree2012}, could be included in the future. Hence, the two-fluid approach could be applied to include collisional damping in the whistler description or the warm Appleton-Hartree equation, as is done in \citeA{Bawaaneh2013}, but studying the effects of Landau or cyclotron damping on whistler waves requires a different framework.

\acknowledgments
This work is supported by funding from the European Research Council (ERC) under the European Unions Horizon 2020 research and innovation programme, Grant agreement No. 833251 PROMINENT ERC-ADG 2018. RK was additionally supported by a joint FWO-NSFC grant G0E9619N and by Internal funds KU Leuven, project C14/19/089 TRACESpace.\\
No datasets were reused or generated for this research.

\bibliography{bibliography}
\end{document}